\documentclass[12pt, preprint]{aastex}

\usepackage{graphicx}

\shorttitle{LMXB Models in Ellipticals: NGC3379 and NGC4278}
\shortauthors{T. Fragos et al.}

\begin{document}

\title{Models for Low-Mass X-Ray Binaries in the Elliptical Galaxies NGC3379 and NGC4278: Comparison with Observations} 
\author{T.\ Fragos$^{1}$, V.\ Kalogera$^{1}$, K.\ Belczynski$^{2}$, G.\ Fabbiano$^{3}$, D.-W.\ Kim$^{3}$, N.\ J.\ Brassington$^{3}$, L.\ Angelini$^{4}$, R.\ L.\ Davies$^{5}$, J.\ S.\ Gallagher$^{6}$, A.\ R.\ King$^{7}$, S.\ Pellegrini$^{8}$, G.\ Trinchieri$^{9}$, 
S.\ E.\ Zepf$^{10}$, A.\ Zezas$^{3}$} 

\altaffiltext{1}{Northwestern University, Department of Physics and Astronomy, 2145 Sheridan Road, Evanston, IL 60208, USA}
\altaffiltext{2}{New Mexico State University, Department of Astronomy, 1320 Frenger Mall, Las Cruces, NM 88003, USA}
\altaffiltext{3}{Harvard-Smithsonian Center for Astrophysics, 60 Garden Street, Cambridge, MA 02138}
\altaffiltext{4}{Laboratory for High Energy Astrophysics, NASA Goddard Space Flight Center, Code 660, Greenbelt, MD 20771}
\altaffiltext{5}{Denys Wilkinson Building, University of Oxford, Keble Road, Oxford OX1 3RH, UK}
\altaffiltext{6}{Astronomy Department, University of Wisconsin, 475 North Charter Street, Madison, WI 53706}
\altaffiltext{7}{University of Leicester, Leicester LE1 7RH, UK}
\altaffiltext{8}{Dipartimento di Astronomia, Universita` di Bologna, Via Ranzani 1, 40127 Bologna, Italy}
\altaffiltext{9}{INAF Observatorio Astronomico di Brera, Via Brera 28, 20121 Milan, Italy}
\altaffiltext{10}{Department of Physics and Astronomy, Michigan State University, East Lansing, MI 48824-2320}

\email{tassosfragos@northwestern.edu, vicky@northwestern.edu, kbelczyn@nmsu.edu,  gfabbiano@cfa.harvard.edu, kim@cfa.harvard.edu, nbrassington@head.cfa.harvard.edu,  angelini@davide.gsfc.nasa.gov, rld@astro.ox.ac.uk,  jsg@astro.wisc.edu, ark@star.le.ac.uk, silvia.pellegrini@unibo.it, ginevra.trinchieri@brera.inaf.it, zepf@pa.msu.edu,  azezas@cfa.harvard.edu}

\begin{abstract}

	We present theoretical models for the formation and evolution of populations of low-mass X-ray binaries (LMXB) in the two elliptical galaxies NGC 3379 and NGC 4278. The models are calculated with the recently updated \emph{StarTrack} code \citep{Belczynski2006}, assuming only a primordial galactic field LMXB population. StarTrack is an advanced population synthesis code that has been tested and calibrated using detailed binary star calculations and incorporates all the important physical processes of binary evolution. The simulations are targeted to modeling and understanding the origin of the X-ray luminosity functions (XLF) of point sources in these galaxies. For the first time we explore the population XLF down to luminosities of $3\times10^{36}\, \rm erg\,s^{-1}$, as probed by the most recent observational results \citep{Kim2006}. We consider models for the formation and evolution of LMXBs in galactic fields with different CE efficiencies, stellar wind prescriptions, magnetic braking laws and initial mass functions. We identify models that produce an XLF in excellent agreement with the observations both in shape and absolute normalization. We also find that the treatment of the outburst luminosity of transient systems remains a crucial factor for the determination of the XLF since the modeled populations are dominated by transient X-ray systems.

\end{abstract}


\section{INTRODUCTION}

%\textbf{talk about diffuse emmission}

A Low Mass X-ray binary (LMXB) is a Roche lobe overflowing, mass-transfering binary system with a compact object accretor, either a black hole (BH) or a neutron star (NS), and a low mass ($\gtrsim 1\,\rm M_{\odot}$) donor.  Since the late 80's it has been suggested that LMXBs should exist in early type galaxies, E and S0, and that they might even dominate the X-ray emission \citep{TF1985, fabbiano1989, KFT1992}. The stellar populations in these galaxies are typically old and homogeneous. Massive stars have already evolved to compact objects and LMXBs are probably the only sources with X-ray luminosities above $10^{36}\, \rm erg\,s^{-1}$. 

Uncontroversial detection of LMXBs in early type galaxies became possible only this last decade with \textit{Chandra's} increased angular resolution \citep{Fabbiano2006, SIB2000}. The spectra of individual X-ray sources are consistent with those expected from LMXB models and the LMXBs observed in the Milky-way and M31 \citep{HB2006, IAB2003}. For many galaxies observed with \textit{Chandra}, the XLFs have been derived and they can usually be fitted with a single or a broken power law. The detections limit for these  surveys is usually a few times $10^{37}\,\rm erg\,s^{-1}$. \citet{KF2004} derived XLFs for 14 early type galaxies and they included completeness corrections. Each XLF is well fitted with a single power law with cumulative slope between -0.8 and -1.2. The composite XLF of these galaxies though is not consistent with a single power law. There is a prominent break at $(5\pm1.6)\times 10^{38}\, \rm erg\,s^{-1}$, close to the Eddington luminosity ($L_{\rm Edd}$) of a helium accreting NS-LMXB. This break might be hidden in the individual XLFs due to poor statistics \citep[see also][]{SIB2000, KMZ2002, Jordan2004, Gilfanov2004}. Other recent studies by \citet{JCBG2003},  \citet{SSI2003} and \citet{Jordan2004} suggested a break of the XLF at a higher luminosity ($\sim 10^{39}\, erg\,s^{-1}$). The exact position and the nature of these breaks are still somewhat controversial, as the correct interpretation of the observed XLFs relies significantly on the proper completeness correction when looking at luminosities close to the detection limit and small number statistics at the high-end of the XLF.  

Recent \textit{Chandra} observations \citep{Kim2006} have yielded the first low-luminosity XLFs of LMXBs for two typical old elliptical galaxies, NGC3379 and NGC4278.  The detection limit in these observations is $\sim 3\times 10^{36}\rm \, erg\, s^{-1}$ which is about an order of magnitude lower than in most previous surveys of early type galaxies. The observed XLFs of the two ellipticals extend only up to $6\times 10^{38}\,\rm erg\, s^{-1}$ and are well represented by a single power law with a slope (in a differential form) of $1.9\pm0.1$.

When \textit{Chandra} observations are compared with optical images from \textit{Hubble} or other ground based telescopes, it is generally found that a significant fraction of the LMXBs are inside globular clusters (GC). On average 4-5\% of the GCs in a given galaxy are associated with a LMXB (with $L_x>\sim 10^{37}\rm \,erg \, s^{-1}$), while the fraction of LMXBs located in GC, varies from 10\% to 70\% depending on the type of the galaxy and its GC specific frequency. At present the origin and the properties of these systems, both in GCs and the field, are not yet well understood. It has been noted that XLFs at high luminosities for each sub-group (GCs and field) do not reveal any differences within the statistics of the samples considered \citep{Fabbiano2006, KimE2006, KMZ2007, Jordan2004, Sarazin2003}. However, in more recent studies, \citep{Fabbiano2007} and \citet{VG2007} independently found that the two XLFs (GCs and field sources) show significant differences at low luminisities (below $10^{37}\rm \, erg\, sec^{-1}$), pointing to a different LMXB formation mechanism in GCs. 

The natural question that arises is whether  (\textit{i}) all LMXBs were formed in GCs through dynamical interactions and some eventually escaped or some GCs dissolved in the field, or (\textit{ii}) field LMXBs were born in situ through binary evolution of primordial binaries. The formation rates associated with these two possibilities are not understood well enough to give accurate predictions and use the relative numbers in the samples. \citet{Juett2005} has shown that the observed relationship between the fraction of LMXBs found in GCs and the GC-specific frequency in early-type galaxies is consistent with the galactic field LMXB population being formed in situ. Similarly, \citet{Irwin2005} compared the summed X-ray luminosity of the LMXBs to the number of GCs in a galaxy; in the case of all LMXBs having formed exclusively in GCs, the two should be directly proportional regardless of where the LMXBs currently reside. Instead he found that the proportionality includes an additive offset implying the existence of a LMXB population unrelated to GCs.

In the past, semi-analytical theoretical models have been introduced for the study  of the LMXB population in early galaxies. \citet{White1998} studied the connection between the star formation rates of normal galaxies, i.e. galaxies without an active nucleus, and the formation rate of LMXBs and millisecond pulsars, assuming that all LMXBs are formed from primordial binaries. Considering a time-dependent star formation rate, they showed that the general relativity timescales relevant to the evolution of primordial binaries to LMXBs and to millisecond pulsars, lead to a significant time delay of the peak in the formation rate of these populations after the peak in the star formation rate. In a followup work \citet{Ghosh2001}, using several updated star formation rate models, calculated the evolution of the X-ray luminosity of galaxies. They found that different star formation models lead to very different X-ray luminosity profiles, so the observed X-ray profiles can be used as probes of the star formation history. Finally they compared their models with the first \textit{Chandra} deep imaging observations, to conclude that these first results were consistent with current star formation models. \citet{Piro2002} argued that the majority of LMXBs in the field of elliptical galaxies have red giant donors feeding a thermally unstable disk and stay in this transient phase for at least 75\% of their life. The very luminous X-ray sources ($L_x>10^{39} \rm \, erg\, s^{-1}$) detected in \textit{Chandra} surveys have been suggested to be X-ray binaries with highly super-Eddington mass inflow near the accreting component. In elliptical galaxies these objects have been suggested by King 2002 to be micro-quasar-like, as these galaxies contain no high mass X-ray binaries \citep{King2002}. More recently  \citet{Ivanova2006} also argued that this bright end of the XLF is most likely dominated by transient LMXBs with BH accretors during outburst and it can be used to derive constraints on the BH mass function in LMXBs; they also showed that the standard assumption of a constant transient duty cycle (DC) across the whole population seems to be inconsistent with current observations. 

Semi-analytical population synthesis (PS) models of LMXBs have also been constructed for late type galaxies. \citet{Wu2001} created a simple birth-death model, in which the lifetimes of the binaries are inversely proportional to their X-ray luminosity, and calculated the XLFs of spiral galaxies. His models reproduce some features, such as the luminosity break in the observed XLFs of spiral galaxies. The position of this break depends on the star formation history of the galaxy, and he suggested that it can be used as a probe of the galaxy's merger history. 

The formation of LMXBs in GCs via dynamical interactions is less well studied, since apart from the binary stellar evolution, one has to also take into account the complex cluster dynamics.  \citet{BD2004} considered a semi-analytical model for accretion from degenerate donors onto NSs in ultracompact binaries and showed that binaries with orbital periods of 8-10 minutes and He or C/O white dwarf (WD) donors of 0.06-0.08 $M_\odot$ naturally provide the primary slope (-0.8 for cumulative form) typically derived from XLFs of elliptical galaxies. Ultra-compact systems are predicted to form in the dense GC environment and have relatively short persistent lifetimes ($<3\times 10^6\, \rm yr$) but they form continuously through dynamical interactions. \citet{Ivanova2007} presented PS studies of compact binaries containing NSs in dense GCs. They used \emph{StarTrack} as their PS modeling tool in addition to a simplified treatment for the dynamical interactions. Their models produced a mixed population of LMXBs with red giant and MS donors, and  ultra-compact X-ray binaries; relative formation rates can be comparable but the different sub-populations have very different lifetimes.

 In this paper we investigate the plausibility of an important contribution to the XLFs of these two galaxies from a primordial galactic field LMXB population using advanced PS simulations. In \S 2 we describe briefly the physics included in our PS code and explain in detail the way we are constructing the modeled XLFs and the treatment of transient LMXBs.  We discuss the results of our simulations in \S 3: the modeled XLFs from different models, a statistical comparison with the observed XLFs of the elliptical galaxies NGC3379 and NGC4278, and an analysis of the dependence of the modeled XLF properties on the PS parameters. Finally in \S 4 we discuss the implication of our findings and the caveats of our methods. 

\section{LMXB Population Models}

For the models presented in this study we focus on  LMXBs formed in the galactic field as products of the evolution of isolated primordial binaries. The standard formation channel \citep{BvdH1991, TvdH2006} involves a primordial binary system with a large mass ratio; the more massive star evolves quickly to the giant branch and the system goes into a common envelope (CE) phase. During this phase, the less massive star, which is still dense and unevolved, orbits inside the envelope of the primary and is assumed to remain intact. The orbit of the system changes dramatically though, as orbital energy is lost due to friction between the the unevolved star and the envelope of the giant. Part of the lost orbital energy is used to expel the envelope of the giant star. The fraction of the lost orbital energy that is used to heat up the envelope of the giant star and finally expel it, defines the CE efficiency factor $\alpha_{\rm CE}$. The CE phase results in a binary system with an unevolved low mass main sequence (MS) star orbiting around the core of the massive star in a tighter orbit. The massive core soon reaches core collapse to form a compact object, either a NS or a BH and the binary orbit is altered due to mass loss and possible supernova kicks. If the binary does not get disrupted or merge in any of the stages described above, angular momentum loss mechanisms, such as magnetic braking, tides and gravitational wave radiation, will shrink further the orbit and the low-mass companion may evolve off the MS; The companion star eventually overflows its Roche lobe, transfering mass onto the compact object and initiating the system's X-ray phase. An alternative formation channel for NS-LMXBs is through accretion induced collapse of a WD accretor into a NS. These systems have generally very low X-ray luminosity, and do not affect the LMXB population in the luminosity range that we are interested in this paper.

\subsection{Synthesis Code: \textit{StarTrack}}

We perform the simulations presented here with \textit{StarTrack} \citep{BKB2002, Belczynski2006}, a advanced PS code that has been tested and calibrated using detailed mass transfer calculations and observations of binary populations, and incorporates all the important physical processes of binary evolution: 
\textit{(i)} The evolution of single stars and non-interacting binary components, from ZAMS to remnant formation, is followed with analytic formulae \citep{HPT2000}. Various wind mass loss rates that vary with the stellar evolutionary stage are incorporated into the code and their effect on stellar evolution is taken into account.
\textit{(ii)} Throughout the course of binary evolution, the changes in all the orbital properties are tracked. A set of four differential equations is numerically integrated, describing the evolution of orbital separation, eccentricity and component spins, which depend on tidal interactions as well as angular momentum losses associated with magnetic braking, gravitational radiation and stellar wind mass losses.  
\textit{(iii)} All types of mass-transfer phases are calculated: stable driven by nuclear evolution or angular momentum loss and thermally or dynamically unstable. Any system entering the Roche lobe over-flow (RLOF) is assumed to become immediately circularized and synchronized. If  dynamical instability is encountered the binary may enter a CE phase. For the modeling of this phase we use the standard energy balance prescription.
\textit{(iv)} The SN explosion is treated taking into account mass-loss as well as SN asymmetries (through natal kicks to NSs and BHs at birth). The distribution of the SN kick magnitudes is inferred from observed velocities of radio pulsars. For this project we use the distribution derived by \citet{Hobbs2005} which is a single Maxwellian with $\sigma=265\rm \, km \, sec^{-1}$. It is however assumed that NS formation via electron capture or accretion induced collapse does not lead to SN kicks.
\textit{(v)} Finally, the X-ray luminosity of accreting binaries with NS and BH primaries (both for wind-fed and RLOF systems) is calculated. For RLOF-fed systems there is a distinction made between persistent and transient (systems that undergo thermal disk instability), while wind-fed systems are always considered as persistent X-ray sources. The mass-transfer is conservative up to the Eddington limit for persistent X-ray binaries, while transients are allowed to have slightly  super-Eddington luminosity (up to $3\times L_{\rm Edd}$) \citep{TCS1997}. In all cases we apply appropriate bolometric corrections ($\eta_{\rm bol}$) to convert the bolometric luminosity to the observed \emph{Chandra} band.
 A much more detailed description of all code elements, treatments of physical processes and implementation is provided in \citet{Belczynski2006}.

\subsection{Model Parameters for NGC3379 and NGC4278}

In this study we focus on trying to understand the XLF characteristics of the two elliptical galaxies NGC3379 and NGC4278, observed with \emph{Chandra} and reported by \citet{Kim2006}. In the development of our models we incorporate our current knowledge about the characteristics of the stellar population in these galaxies (see Table \ref{galaxyparam}). The observationally determined parameters of the stellar populations, such as their age and metallicity , or their total stellar mass, are similar for NGC3379 and NGC4278. This allows us to develop the same models in our simulations for both of them.  \citet{TF2002} estimated the ages and metallicities of 150 elliptical and late type spiral galaxies using published hight quality spectral line indices. For NGC3379 they are reporting an age of 9.3\, Gyr and a metallicity of [Fe/H]=0.16, while for NGC4278 the corresponding values are 10.7\, Gyr and [Fe/H]=0.14. The two galaxies have very similar optical luminosity and assuming the same light to mass ratio, they should also have similar masses. \citet{Cappellari2006} used I-band observations from the \emph{Hubble Space Telescope} to calculate the total stellar mass of the two galaxies and they found them to be $8.6\times 10^{10} \rm M_{\odot}$ and $9.4\times 10^{10} \rm M_{\odot}$ for N3379 and N4278 respectively. The ratio of the integrated LMXB X-ray luminosity to the optical luminosity is 4 times smaller for NGC3379 which also has six times lower GC specific frequency compared to NGC4278 \citep[see ][]{Kim2006,AZ1998}.

\begin{deluxetable}{lcccc}
\tablecolumns{4}
\tabletypesize{\scriptsize}
\tablewidth{0pt} 
\tablecaption{Galaxy Properties 
\label{galaxyparam}} 
\tablehead{ \colhead{Parameter} & 
	 \colhead{} & 
     \colhead{NGC3379} & 
     \colhead{NGC4278} &
	 \colhead{References}
     }
\startdata
Distance (Mpc)						& &	10.57				&	16.07				& \citet{Tonry2001}\\
Age (Gyr)							& &	9.3					&	10.7				& \citet{TF2002}\\
Metallicity	([Fe/H])				& &	0.16				&	0.14				& \citet{TF2002}\\
Mass	($\rm M_{\odot}$)	& &	$8.6\times 10^{10}$	&	$9.4\times 10^{10}$	& \citet{Cappellari2006}\\
GC Specific Frequency				& & 1.2					&	6.9					& \citet{AZ1998} \\
\enddata

\end{deluxetable}

There is however a number of parameters in our models, for which we do not have any direct guidance from observations. We have no information about the star formation history of the two galaxies and thus we assume a $\delta$-function like star formation episode at time t=0. Unknown are also the Initial mass function (IMF) and the distributions of orbital separation and eccentricity for the primordial binary systems. We adopt two different initial mass functions: Scalo/Kroupa and Salpeter, while  for the distributions of the orbital properties we follow the standard assumptions described in \citet{BKB2002}. Other parameters that can affect the final LMXB population are the binary fraction of the host galaxy, the magnetic braking law adopted and the CE efficiency ($\alpha_{\rm CE}$)\footnote{In our calculations, we combine $\alpha_{\rm CE}$ and $\lambda$ into one CE parameter, where $\lambda$ is a measure of the central concentration of the donor. In the rest of the text, whenever we mention the CE efficiency $\alpha_{\rm CE}$, we practically refer to the product  $\alpha_{\rm CE} \times \lambda$ (see \citet{Belczynski2006} for details.) }. The specific parameters we used to model the ellipticals NGC3379 and NGC4278 are listed in Table \ref{modelparam}.

\begin{deluxetable}{lcc} 
\tablecolumns{3}
\tabletypesize{\scriptsize}
\tablewidth{0pt}
\tablecaption{Model Parameters for NGC3379 and NGC4278 
\label{modelparam}} 
\tablehead{ \colhead{Parameter} & 
     \colhead{Notation} & 
     \colhead{Value} 
     }
\startdata
Star Formation                    &           & $\delta-$function at $t=0$ \\
Population Age          &            &  $9-10\, \rm Gyr$ \\
Metallicity           & $Z$           & 0.03 \\
Total Stellar Mass        & $M_{*}$ &  $ 9 \times 10^{10}\, \rm M{\odot}$ \\
Binary Fraction  & $F_{\rm bin}$ & 50\% \\
IMF &      & Scalo/Kroupa or Salpeter \\
CE Efficiency & $\alpha_{\rm CE}$  &  20\% --  100\% \\
Magnetic Braking &		& \mbox{\citet{RVJ1983} or \citet{IT2003}}

\enddata

\end{deluxetable}

\subsection{Models for the X-ray Luminosity Function}

In our models we keep track of all the binary properties, including the mass-transfer rate ($\dot{M}$), as a function of time for populations of accreting NS and BH. We use the mass-transfer rates to identify the persistent and transient sources in our simulation results. Binaries for mass transfer rate higher than the critical rate $\dot{M}_{\rm crit}$ for the thermal disk instability \citep{DLHC1999,MPH2002}, are considered persistent sources and their X-ray luminosity ($L_{\rm x}$) is calculated directly from the mass transfer rate as 
$$L_{\rm x}=\eta_{\rm bol}\epsilon\frac{GM_{\rm a}\dot{M}}{R_{\rm a}},$$
where the radius of the accretor ($R_{\rm a}$) is 10\, km for a NS and 3 Schwarzschild radii for a BH, $\epsilon$ gives a conversion efficiency of gravitational binding energy to radiation associated with accretion onto a NS (surface accretion $\epsilon=1.0$) and onto a BH (disk accretion $\epsilon=0.5$), and $\eta_{\rm bol}$ is a factor that converts the bolometric luminosity to the X-ray luminosity in the \emph{Chandra} energy band (0.3 - 8 keV). For RLOF accreting BH this conversion factor is estimated to be $\eta_{\rm bol}=0.8$ \citep{Miller2001} while for RLOF accreting NS $\eta_{\rm bol}=0.55$ \citep{Disalvo2002,MC2003,PZDM2004}. The two correction factors $\epsilon$ and $\eta_{\rm bol}$ are applied in both persistent sources and transient sources in outburst.
 
In the context of the thermal disk instability model, mass transferring binaries with $ \dot{M}  < \dot{M}_{\rm crit}$ are considered transient sources, meaning that they spend most of their life in a quiescent state ($T_{\rm quiscent}$), at which they are too faint to be detectable, and they occasionally go into an outburst. The fraction of the time that these systems are in outburst ($T_{\rm outburst}$) defines their DC: 
\begin{equation}{\rm DC} \equiv \frac{T_{\rm outburst}}{T_{\rm outburst}+T_{\rm quiescent}}. \label{dc_def}\end{equation}
Observations of Galactic LMXBs show that transient systems spend most of their life in the quiescent state, hinting at a DC below 20\% \citep{TS1996}. 

The outburst luminosity and the DC of transient is not well understood and cannot be calculated from first principles. Instead we have to rely primarily on empirical constraints and simple theoretical arguments. In our analysis we consider a number of different treatments of these parameters for transients, which we describe in what follows. 

As a first approximation there has been suggested that transient LMXBs emit at their Eddington luminosity ($L_{\rm Edd}$) when they are in outburst. In a different approach \citet{PDM2005} derived an empirical correlation  between the outburst luminosity of Milky Way transient LMXBs with BH accretors and their orbital period P: 

 \begin{equation} L_{\rm x} = \eta_{\rm bol}\epsilon\times \min{\left[2\times L_{\rm Edd} , 2\times L_{\rm Edd} \left(\frac{P}{10 \rm h}\right)\right]} \label{Lx_P}\end{equation}
We can generalize this relation to all transient LMXBs in galaxies other than our own, but we note that there has not been any observational work that shows that NS-LMXBs follow a similar trend. 

A more physical treatment is to assume that in the quiescent state the compact object does not accrete (or accretes an insignificant amount of mass) and matter from the donor is accumulated in the disk. In the outburst state all this matter is accreted onto the compact object emptying again the disk. Taking into account also that the X-ray luminosity probably cannot exceed $L_{\rm Edd}$ by more than a factor of 2 \citep[cf.][]{TCS1997}, we end up with a definition of the outburst luminosity as:

 \begin{equation} L_{\rm x} = \eta_{\rm bol}\epsilon\times
 \min{\left(3\times L_{\rm Edd} , \frac{GM_{\rm a}\dot{M}_{\rm d}}{R_{\rm a}} \times \frac{1}{\rm DC}\right)} \label{Lx_M} \end{equation}

In the equation above, DC is unknown. \cite{DLM2006} studied accretion disk models for cataclysmic variables that are thought to experience the same thermal disk instability (dwarf novae). They found a correlation between the DC of the system and the rate at which the donor star is losing mass $\dot{M}_{\rm d}$. The exact relation of these to quantities depends on the values of the disk's viscosity parameters, but the general behavior can be approximated by:
\begin{equation} DC=\left(\frac{\dot{M}_{\rm d}}{\dot{M}_{\rm crit}} \right)^{2}. \label{dc} \end{equation}
Plugging eq.(\ref{dc}) into eq.(\ref{Lx_M}) we eliminate the DC dependence and get an expression for the outburst luminosity of a transient system that depends only quantities which are directly calculated in our population modeling:
 \begin{equation}  L_{\rm x} = \eta_{\rm bol}\epsilon\times \min{\left(3\times L_{\rm Edd} , \frac{GM_{\rm a}\dot{M}_{\rm crit}^{2}}{R_{\rm a}\dot{M}_{\rm d}} \right)} \label{Lx_phys} \end{equation}

The accretion disk models by \cite{DLM2006} assume accretion onto a compact object with a hard surface and it is not obvious that the same results will apply for accretion onto a BH. In order to take into account all the available information, empirical and theoretical, about LMXB transient behavior, we treat  BH and NS-LMXBs differently and define the outburst luminosity as:  
\begin{equation}
L_{\rm x}=\eta_{\rm bol}\epsilon\times \Bigg\{
\begin{array}{ll}
\min{\left[2\times L_{\rm Edd} , 3\times L_{\rm Edd} \left(\frac{P}{10 \rm h}\right)\right]}, &  \rm{for\ BH\ acc.} \\
& \\
\min{\left(2\times L_{\rm Edd} , \frac{GM_{\rm a}\dot{M}_{\rm crit}^{2}}{R_{\rm a}\dot{M}_{\rm crit}} \right)}, & \rm{for\ NS\ acc.}\\
\end{array}
\label{Lx_mix}
\end{equation}

We note that for BH-LMXBs, we adopt a single DC value for simplicity and lack of other information, although we have no clear physical reason to believe that all BH systems have the same DC. It is believed that the DC of BH systems is smaller than that of NS systems and on the order of $5\%$ \citep{TS1996}. We found that in all our models NS accretors greatly outnumber BH accretors. BH systems only have an important contribution at high luminosities, where the error bars in the observed XLFs \citep[see ][]{Kim2006} are too large to give us any tight constraints for our models. As we will see in section 3.1, this treatment eq.(\ref{Lx_mix}) gives us the best agreement with observations.

To construct the XLF we consider a snapshot of the whole population at the time we are interested in and we identify the LMXBs as transient or persistent. If a system is transient we decide whether it is in outburst or in quiescence according to its DC and either assign an outburst luminosity or discard the system as quiescent and hence too faint to contribute to the XLF. We then construct the XLF  by calculating the cumulative X-ray luminosity distribution of the sources that are detectable (persistent and transient in outburst). We note that the age of the elliptical galaxies NGC3379 and NGC4278 and hence their LMXB population is known only to within $\sim 1$\,Gyr. Consequently we cannot just choose a unique snapshot of the population. Instead we construct the XLF by considering the time window of 9 - 10\,Gyr  and considering time slices separated by 1\,Myr. We construct the XLF at each of these timeslices and we take the average to represent the XLF that corresponds to the time window of interest.  Doing so, we also improve the statistics of our model sample. 

It is computationally impossible to evolve enough binaries that would correspond to the total initial number of binaries in an elliptical galaxy ($\sim 10^9$ binaries). For each model we evolve $10^6$ binaries which takes about two months of CPU time on a modern processor. We then normalize to the total mass of the galaxy in question, taking into account the initial binary fraction and the initial mass function.

\section{Results}

\subsection{Exploring the parameter space}

\begin{deluxetable}{cccc}
\tablecolumns{4}
\tabletypesize{\scriptsize}
\tablewidth{0pt}
\tablecaption{Population Synthesis Models: For each of the models listed below we applied two magnetic braking law prescriptions. In the rest of the text the \citet{IT2003} prescription will be denoted with the ``IT'' superscript at the end of the model name, while the \citet{RVJ1983} prescription with the ``RVJ'' superscript.
\label{models}} 
\tablehead{ \colhead{Model} & 
     \colhead{$\alpha_{\rm CE}$} & 
     \colhead{\ \ \ \ IMF \ \ \ \ } & 
     \colhead{$\eta_{\rm wind}$} 
     }
\startdata
1	& 0.2	& Salpeter		& 0.25	\\
2	& 0.2	& Scalo/Kroupa	& 0.25	\\
3	& 0.2	& Salpeter		& 1.0	\\
4	& 0.2	& Scalo/Kroupa	& 1.0	\\
5	& 0.3	& Salpeter		& 0.25	\\
6	& 0.3	& Scalo/Kroupa	& 0.25	\\
7	& 0.3	& Salpeter		& 1.0	\\
8	& 0.3	& Scalo/Kroupa	& 1.0	\\
9	& 0.4	& Salpeter		& 0.25	\\
10	& 0.4	& Scalo/Kroupa	& 0.25	\\
11	& 0.4	& Salpeter		& 1.0	\\
12	& 0.4	& Scalo/Kroupa	& 1.0	\\
13	& 0.5	& Salpeter		& 0.25	\\
14	& 0.5	& Scalo/Kroupa	& 0.25	\\
15	& 0.5	& Salpeter		& 1.0	\\
16	& 0.5	& Scalo/Kroupa	& 1.0	\\
17	& 0.6	& Salpeter		& 0.25	\\
18	& 0.6	& Scalo/Kroupa	& 0.25	\\
19	& 0.6	& Salpeter		& 1.0	\\
20	& 0.6	& Scalo/Kroupa	& 1.0	\\
21	& 0.7	& Salpeter		& 0.25	\\
22	& 0.7	& Scalo/Kroupa	& 0.25	\\
23	& 0.7	& Salpeter		& 1.0	\\
24	& 0.7	& Scalo/Kroupa	& 1.0	\\
25	& 1.0	& Salpeter		& 0.25	\\
26	& 1.0	& Scalo/Kroupa	& 0.25	\\
27	& 1.0	& Salpeter		& 1.0	\\
28	& 1.0	& Scalo/Kroupa	& 1.0	\\

\enddata

\end{deluxetable}

\begin{deluxetable}{ccccc}
\tablecolumns{5}
\tabletypesize{\scriptsize}
\tablewidth{0pt}
\tablecaption{Treatment of Transient LMXBs: We tried 6 different prescriptions for the determination of the DC and outburst luminosity of transient LMXBs. 
\label{models_tran}} 
\tablehead{ \colhead{\textbf{Model Name}} & 
     \colhead{$L_{\rm x,NS}$} &
     \colhead{$L_{\rm x,BH}$} &
     \colhead{$DC_{\rm NS}$} &
     \colhead{$DC_{\rm BH}$} 
     }
\startdata
A	& eq.(\ref{Lx_mix})	& eq.(\ref{Lx_mix})	& eq.(\ref{dc})	&	5\%  	\\
B	& eq.(\ref{Lx_M})	& eq.(\ref{Lx_M})	& 1\%		 	&	1\%		\\
C	& eq.(\ref{Lx_M})	& eq.(\ref{Lx_M})	& 7\%		 	&	7\%		\\
D	& eq.(\ref{Lx_M})	& eq.(\ref{Lx_M})	& 15\%		 	&	15\%	\\
E	& $L_{\rm Edd}$			& $L_{\rm Edd}$			& 10\%			&	10\%	\\
F	& eq.(\ref{Lx_P})	& eq.(\ref{Lx_P})	& 10\%			&	10\%  	\\

\enddata

\end{deluxetable}

One of the implicit weaknesses of PS models is the large number of free parameters that one can vary and fine tune in order to get the desirable result.  
There are physical processes involved in the evolution of a binary system, such as stellar winds and magnetic braking, which are not fully understood. In this case various prescriptions are typically used to model them. Fortunately: (i)  the result of interest to us in this study (XLF) is not sensitive to all of these model parameters; (ii) we can use some empirical knowledge from observations to constrain these parameters.  We study a total of 336 models. In Table \ref{models} we list 28 combinations of PS input parameters we study (CE efficiency, initial mass function, wind strength) and in Table \ref{models_tran} we list the 6 different prescriptions we use for the determination of the DC and outburst luminosity of transient LMXBs. For each of the parameter combination from Tables \ref{models} and \ref{models_tran}, we try two prescriptions for the magnetic braking law, by \citet{IT2003} and \citet{RVJ1983}. Parameters not mentioned here are set as in the standard model considered in \citet{Belczynski2006}. We name our models using a combination of a number from Table \ref{models} which denotes the PS parameters of the models, a letter from Table \ref{models_tran} which denotes the prescription we use for the treatment of transient LMXBs and a superscript which denotes the magnetic braking law used.

\begin{figure}
\plotone{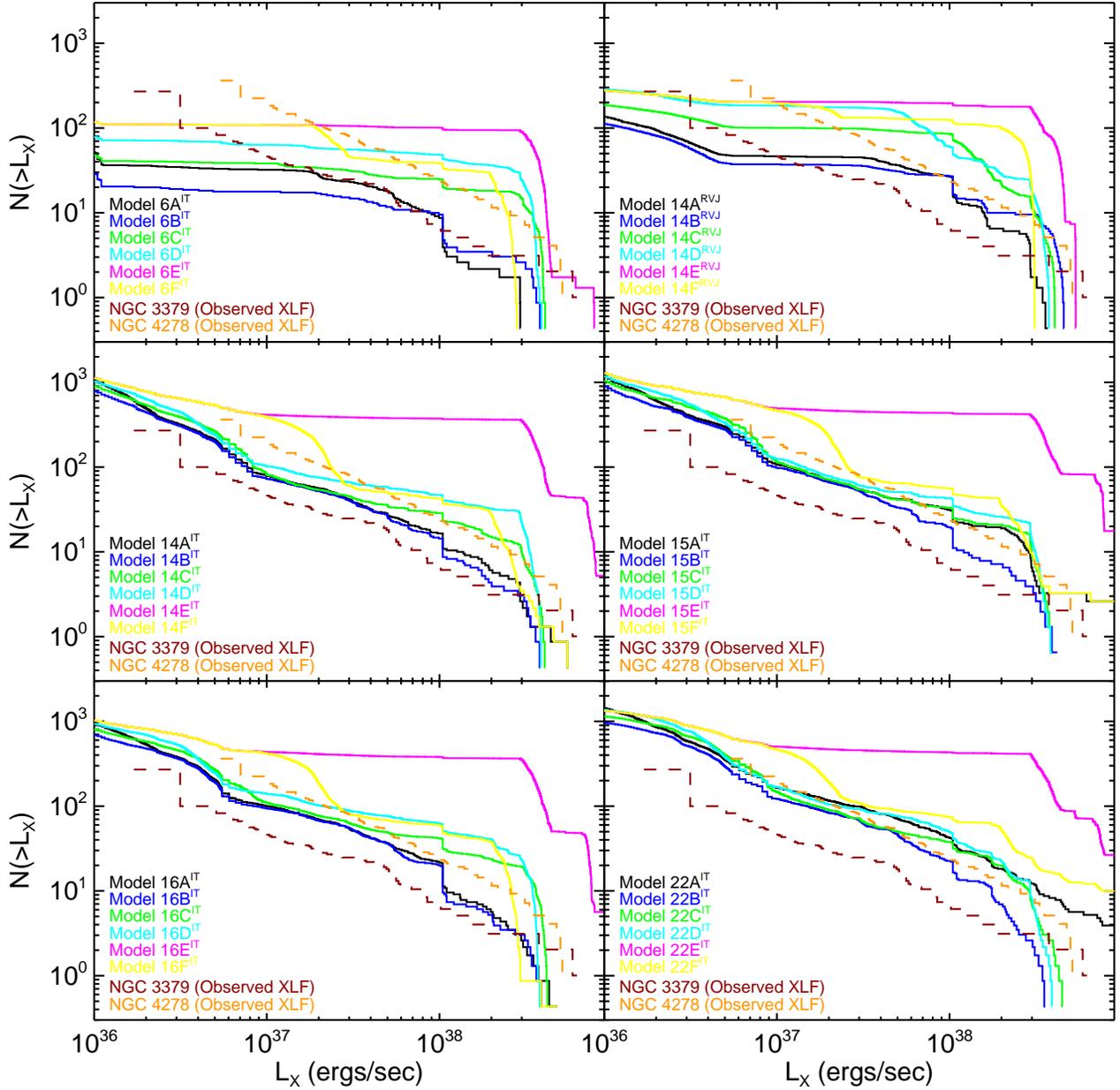}
\caption{Model XLFs for 36 selected LMXB population. In each panel all the PS parameters are kept constant, and only the modeling of transient systems changes. For comparison the observed XLFs of NGC3379 (dark red) and of NGC4278 (orange) are drawn. The modeling of the transient systems and their outburst characteristics can be more important than the usual PS parameters; Commonly used assumptions such as  assigning the outburst luminosity of the transient LMXBs to be equal to $L_{\rm Edd}$ (transient treatment E) lead to XLFs clearly inconsistent with the observations.}
\label{xlf_dc}
\end{figure}

The analysis of several population simulations enables us to identify general behaviors of how the  LMXB population is affected by changes to different parameters. In Figure  \ref{xlf_dc} we show the modeled XLFs for 36 selected models, separated in 6 panels. In each panels all the parameters that determine the formation and evolution of LMXBs are kept constant, and we just change the modeling of transient systems (their DC and outburst luminosity). The observed XLFs of the two galaxies, NGC3379 and NGC4278, are also plotted for comparison. 

We find that the stellar wind strength  alters significantly only the BH-LMXB population and thus only the high luminosity region of the modeled XLF. Stellar winds are important mainly in the evolution of massive stars which, depending on the wind prescription assumed, may lose a significant part of their envelope. Weaker stellar winds result in more massive pre-supernova cores and thus to the formation of more BHs. On the other hand, the initial mass function affects the LMXB population globally. A flatter (Salpeter) initial mass function increases the overall number of sources with luminosities above $10^{37}\, \rm erg\,s^{-1}$ but still gives results consistent with the observations.  The modeled XLF is more sensitive to the CE efficiency $\alpha_{\rm CE}$. Smaller $\alpha_{\rm CE}$ values lead to more mergers among LMXB progenitors and therefore  decrease the overall formation rate of LMXBs. If the CE efficiency is set to a very small value ($\alpha_{\rm CE}<0.2$), it becomes very difficult to form any LMXB at all \citep{KB1998}, while a high CE efficiency $\alpha_{\rm CE}>0.6$, clearly overproduce LMXBs. It is also clear that the adopted magnetic braking law greatly affects the characteristics of the model LMXB population. The XLFs of all 336 models can be seen in the online material supplemental to this paper.

The ratio of persistent to transient LMXBs produced from our models, including systems both in outburst and in quiescent, is on the order of 1:20. So even if transient systems have a small DC ($DC<10\%$), their contribution to the shape of the XLF is important and some times they may even dominate the population \citet[as discussed by ][]{Piro2002}. From this simple argument, we can understand that the shape of the modeled XLFs  should be very sensitive to  the DC and the outburst luminosity of transient systems. Figure \ref{xlf_dc} confirms this last statement, as we see that the larger variation in the shape of the modeled XLFs comes from different prescriptions in the treatment of transient systems and not from changing the binary evolution parameters.

\begin{deluxetable}{ccccccccc}
\tablecolumns{1}
\tabletypesize{\scriptsize}
\tablewidth{0pt} 
\tablecaption{Statistical comparison of the models shown in Figure (\ref{xlf_dc}) to the observed XLFs of NGC3379 and NGC4278. We used four different statistical tests for each case (Kolmogorov-Smirnov test, Kuiper test, Wilcoxon rank-sum test and $\chi^2$ goodness of fit test) and we are listing the probabilities, derived from each test, that a modeled XLF is consistent with the observed ones. For the statistical comparison of the complete list of models please see the electronic supplemental material.
\label{stat_test}} 
\tablehead{ \colhead{Model} & 
			\multicolumn{4}{c}{NGC 3379} &
			\multicolumn{4}{c}{NGC 4278} \\
	 \colhead{}	&	
     \colhead{$P_{\rm KS}$} & 
     \colhead{$P_{\rm Kuiper}$} & 
     \colhead{$P_{\rm \chi^2}$} &
     \colhead{$P_{\rm RS}$} & 
     \colhead{$P_{\rm KS}$} & 
     \colhead{$P_{\rm Kuiper}$} & 
     \colhead{$P_{\rm \chi^2}$} &
     \colhead{$P_{\rm RS}$}  

}
\startdata

\mbox{$\rm 6A^{IT}$}    &    0.037    &    0.065    &    0.006    &    0.022    &    \textbf{0.275}    &    0.020    &    \textbf{0.202}    &    0.046    \\
\mbox{$\rm 6B^{IT}$}    &    2.18e-04    &    0.002    &    5.63e-05    &    1.53e-04    &    0.002    &    0.007    &    0.012    &    3.92e-04    \\
\mbox{$\rm 6C^{IT}$}    &    3.89e-08    &    1.10e-07    &    7.75e-07    &    1.10e-06    &    4.89e-09    &    7.27e-09    &    5.96e-08    &    1.04e-09    \\
\mbox{$\rm 6D^{IT}$}    &    1.61e-13    &    1.68e-13    &    0.00e+00    &    5.48e-11    &    4.67e-15    &    1.12e-15    &    0.00e+00    &    1.33e-15    \\
\mbox{$\rm 6E^{IT}$}    &    1.40e-24    &    7.98e-25    &    0.00e+00    &    0.00e+00    &    5.03e-34    &    1.31e-35    &    0.00e+00    &    0.00e+00    \\
\mbox{$\rm 6F^{IT}$}    &    0.001    &    1.60e-05    &    0.011    &    0.00e+00    &    6.60e-13    &    1.97e-16    &    0.00e+00    &    7.46e-11    \\
\mbox{$\rm 14A^{RVJ}$}    &    2.77e-07    &    6.89e-07    &    0.00e+00    &    4.96e-10    &    3.34e-04    &    1.30e-04    &    0.003    &    6.39e-04    \\
\mbox{$\rm 14B^{RVJ}$}    &    1.34e-10    &    3.33e-10    &    0.00e+00    &    2.93e-10    &    2.60e-08    &    3.35e-08    &    5.13e-06    &    4.68e-07    \\
\mbox{$\rm 14C^{RVJ}$}    &    4.94e-19    &    1.83e-19    &    0.00e+00    &    0.00e+00    &    3.61e-18    &    2.11e-19    &    0.00e+00    &    0.00e+00    \\
\mbox{$\rm 14D^{RVJ}$}    &    6.87e-10    &    5.52e-10    &    0.00e+00    &    0.00e+00    &    6.96e-06    &    1.25e-06    &    4.06e-04    &    1.03e-10    \\
\mbox{$\rm 14E^{RVJ}$}    &    6.72e-27    &    6.11e-28    &    0.00e+00    &    0.00e+00    &    5.05e-37    &    2.14e-35    &    0.00e+00    &    0.00e+00    \\
\mbox{$\rm 14F^{RVJ}$}    &    4.38e-10    &    3.35e-10    &    2.38e-07    &    0.00e+00    &    3.17e-24    &    3.97e-29    &    0.00e+00    &    0.00e+00    \\
\mbox{$\rm 14A^{IT}$}    &    \textbf{0.804}    &    \textbf{0.553}    &    \textbf{0.351}    &    \textbf{0.528}    &    \textbf{0.661}    &    \textbf{0.654}    &    \textbf{0.302}    &    \textbf{0.438}    \\
\mbox{$\rm 14B^{IT}$}    &    \textbf{0.780}    &    \textbf{0.785}    &    \textbf{0.291}    &    \textbf{0.536}    &    \textbf{0.890}    &    \textbf{0.779}    &    \textbf{0.462}    &    \textbf{0.449}    \\
\mbox{$\rm 14C^{IT}$}    &    0.049    &    0.045    &    \textbf{0.171}    &    0.018    &    3.24e-05    &    4.32e-05    &    3.29e-04    &    3.36e-04    \\
\mbox{$\rm 14D^{IT}$}    &    1.70e-04    &    3.22e-04    &    0.001    &    0.002    &    1.14e-08    &    5.67e-09    &    0.00e+00    &    2.76e-08    \\
\mbox{$\rm 14E^{IT}$}    &    3.82e-28    &    3.38e-25    &    0.00e+00    &    0.00e+00    &    0.00e+00    &    0.00e+00    &    0.00e+00    &    0.00e+00    \\
\mbox{$\rm 14F^{IT}$}    &    1.95e-06    &    5.16e-06    &    0.001    &    0.00e+00    &    2.21e-08    &    3.94e-09    &    1.31e-06    &    4.57e-07    \\
\mbox{$\rm 15A^{IT}$}    &    \textbf{0.193}    &    \textbf{0.303}    &    \textbf{0.199}    &    0.065    &    0.001    &    0.004    &    0.002    &    2.36e-04    \\
\mbox{$\rm 15B^{IT}$}    &    \textbf{0.871}    &    \textbf{0.647}    &    \textbf{0.366}    &    \textbf{0.444}    &    \textbf{0.837}    &    \textbf{0.509}    &    \textbf{0.483}    &    0.066    \\
\mbox{$\rm 15C^{IT}$}    &    \textbf{0.298}    &    \textbf{0.151}    &    \textbf{0.454}    &    0.035    &    2.92e-04    &    5.57e-04    &    9.97e-04    &    5.96e-05    \\
\mbox{$\rm 15D^{IT}$}    &    0.052    &    0.042    &    \textbf{0.194}    &    0.010    &    2.04e-07    &    2.11e-07    &    2.86e-06    &    2.87e-07    \\
\mbox{$\rm 15E^{IT}$}    &    1.72e-26    &    1.51e-23    &    0.00e+00    &    0.00e+00    &    1.96e-44    &    2.61e-41    &    0.00e+00    &    0.00e+00    \\
\mbox{$\rm 15F^{IT}$}    &    4.27e-06    &    1.28e-06    &    0.005    &    1.11e-16    &    4.95e-09    &    2.47e-08    &    1.19e-07    &    1.67e-07    \\
\mbox{$\rm 16A^{IT}$}    &    \textbf{0.832}    &    \textbf{0.778}    &    \textbf{0.294}    &    \textbf{0.397}    &    \textbf{0.498}    &    \textbf{0.121}    &    \textbf{0.377}    &    0.022    \\
\mbox{$\rm 16B^{IT}$}    &    \textbf{0.585}    &    \textbf{0.629}    &    \textbf{0.204}    &    \textbf{0.471}    &    \textbf{0.276}    &    \textbf{0.129}    &    \textbf{0.328}    &    0.007    \\
\mbox{$\rm 16C^{IT}$}    &    0.009    &    0.018    &    0.044    &    0.004    &    5.22e-07    &    3.61e-07    &    3.91e-05    &    6.95e-07    \\
\mbox{$\rm 16D^{IT}$}    &    1.91e-04    &    3.26e-04    &    0.001    &    0.003    &    1.54e-09    &    1.16e-10    &    0.00e+00    &    8.82e-08    \\
\mbox{$\rm 16E^{IT}$}    &    8.10e-26    &    5.55e-23    &    0.00e+00    &    0.00e+00    &    6.45e-44    &    3.10e-41    &    0.00e+00    &    0.00e+00    \\
\mbox{$\rm 16F^{IT}$}    &    2.78e-05    &    9.86e-06    &    0.008    &    4.44e-16    &    1.43e-10    &    1.57e-13    &    1.79e-07    &    3.03e-09    \\
\mbox{$\rm 22A^{IT}$}    &    \textbf{0.184}    &    \textbf{0.566}    &    0.095    &    \textbf{0.606}    &    0.045    &    \textbf{0.103}    &    0.035    &    \textbf{0.358}    \\
\mbox{$\rm 22B^{IT}$}    &    \textbf{0.887}    &    \textbf{0.921}    &    \textbf{0.331}    &    \textbf{0.793}    &    \textbf{0.845}    &    \textbf{0.575}    &    \textbf{0.442}    &    \textbf{0.426}    \\
\mbox{$\rm 22C^{IT}$}    &    \textbf{0.424}    &    \textbf{0.122}    &    \textbf{0.408}    &    0.046    &    0.003    &    7.28e-04    &    0.025    &    0.019    \\
\mbox{$\rm 22D^{IT}$}    &    0.041    &    0.020    &    \textbf{0.160}    &    0.090    &    3.17e-05    &    9.40e-06    &    2.01e-04    &    6.87e-04    \\
\mbox{$\rm 22E^{IT}$}    &    6.17e-25    &    2.35e-22    &    0.00e+00    &    0.00e+00    &    3.48e-43    &    2.40e-40    &    0.00e+00    &    0.00e+00    \\
\mbox{$\rm 22F^{IT}$}    &    2.66e-04    &    6.07e-05    &    0.005    &    2.66e-11    &    1.46e-09    &    1.13e-08    &    0.00e+00    &    4.51e-06    \\

\enddata

\end{deluxetable}

%perform a linear regression (in log-log) and then compare the exponents. 

We compare quantitatively our modeled XLFs to the observed ones from the two elliptical galaxies NGC3379 and NNGC4278, using statistical tests that are not making an a priori assumption about the functional form of the observed XLF, and we report the probability that the two data sets are drawn from the same distribution. In order to make a fair comparison between models and observations, we include in our statistical analysis only systems with luminosity above the completeness limit for the specific observations ( $>10^{37}\, \rm erg\,s^{-1}$ for NGC3379 and  $>3\times 10^{37}\, \rm erg\,s^{-1}$ for NGC4278). We have used three hypothesis testing statistical methods, the widely used Kolmogorov-Smirnov (KS), the Kuiper test which is a variation of the KS, equally sensitive in the whole range of the XLF and not only in the mid-range and the Wilcoxon rank-sum test (RS). We have also used the $\chi^2$ goodness of fit test \citep{NR2007}. We note here that all four statistical tests are comparing only the shape of the XLFs and not their absolute normalization.

In Table \ref{stat_test} we list the P values (probability that the two data sets are drawn from the same distribution) for all four tests, comparing the models shown in Figure \ref{xlf_dc} with each of the two observed XLFs from NGC3379 and NGC4278. For the statistical analysis of the complete list of models examined in this work, please see the online supplemental material. We use this analysis to draw general conclusions regarding the behavior of our models. In all cases, when we assume the outburst luminosity of the transient LMXBs to be equal to $L_{\rm Edd}$ or use eq.(\ref{Lx_P}) (transient treatment E and F, see Table \ref{models_tran}), the P values are extremely low and we can confidently say that this treatment of transient systems is highly unlikely to be correct. Assigning the same constant DC to all transients and then calculating the outburst luminosity using eq.(\ref{Lx_M}), sometimes leads to results consistent with the observations depending on the rest of the parameters used in the models. When this constant DC is as low as 1\% (transient treatment B), the modeled XLFs resemble remarkably the observed ones. This happens because such a low DC practically removes the contribution of transient LMXBs to the total XLF. However, a DC of 1\% is unrealistically low based on both observational evidence and theoretical predictions. Finally, the transient treatment A consistently provides the best agreement with observations, and in addition it is the one most physically motivated. One of our primary conclusions from this analysis is that careful modeling of the transient systems and their outburst characteristics is very important, more important than the usual PS parameters. 

Studying the complete list of models (see online supplemental material) becomes evident that the magnetic braking law applied during the evolution of the LMXBs drastically changes the resulted population. Models where the \citet{RVJ1983} prescription is used, produce XLFs inconsistent with the observations. The \citet{RVJ1983} braking law prescription predicts much stronger angular momentum losses, compared to the \citet{IT2003} one. Very strong angular momentum loss due to magnetic breaking lead to population of LMXBs where only wide binaries have avoided a possible merger. Our findings in favor of a milder magnetic braking law \citep{IT2003} are in agreement with earlier work by \citet[][ and references there in]{VKK1998}.  

Our statistical analysis tests only the shape of the XLF. However, within the uncertainties of the total mass of the two galaxies most of models we consider turn out to have the right normalization, and therefore this is not a strongly discriminating constraint. We note also that the PS model normalization can be very sensitive to certain binary evolution parameters, such as the distribution of the mass ratios between the initial masses of the two binary components. Without a proper multi-dimensional coverage of the parameter space and given the uncertainties in the total galaxy mass, we cannot use the total number of LMXBs observed as a formal constraint. Instead, we remain satisfied that most of the models we consider give us a total number of LMXBs, in the observed luminosity regime, with a factor of 3 from the observed sample. This factor is comparable to the galaxy mass uncertainty \citep{Cappellari2006}. 

In the discussion that follows we are using model $\rm 14A^{IT}$ as our \emph{standard} model that gives an XLF closest both in shape and normalization to the observed ones.

\subsection{Analyzing the LMXB population}

\begin{figure}
\plottwo{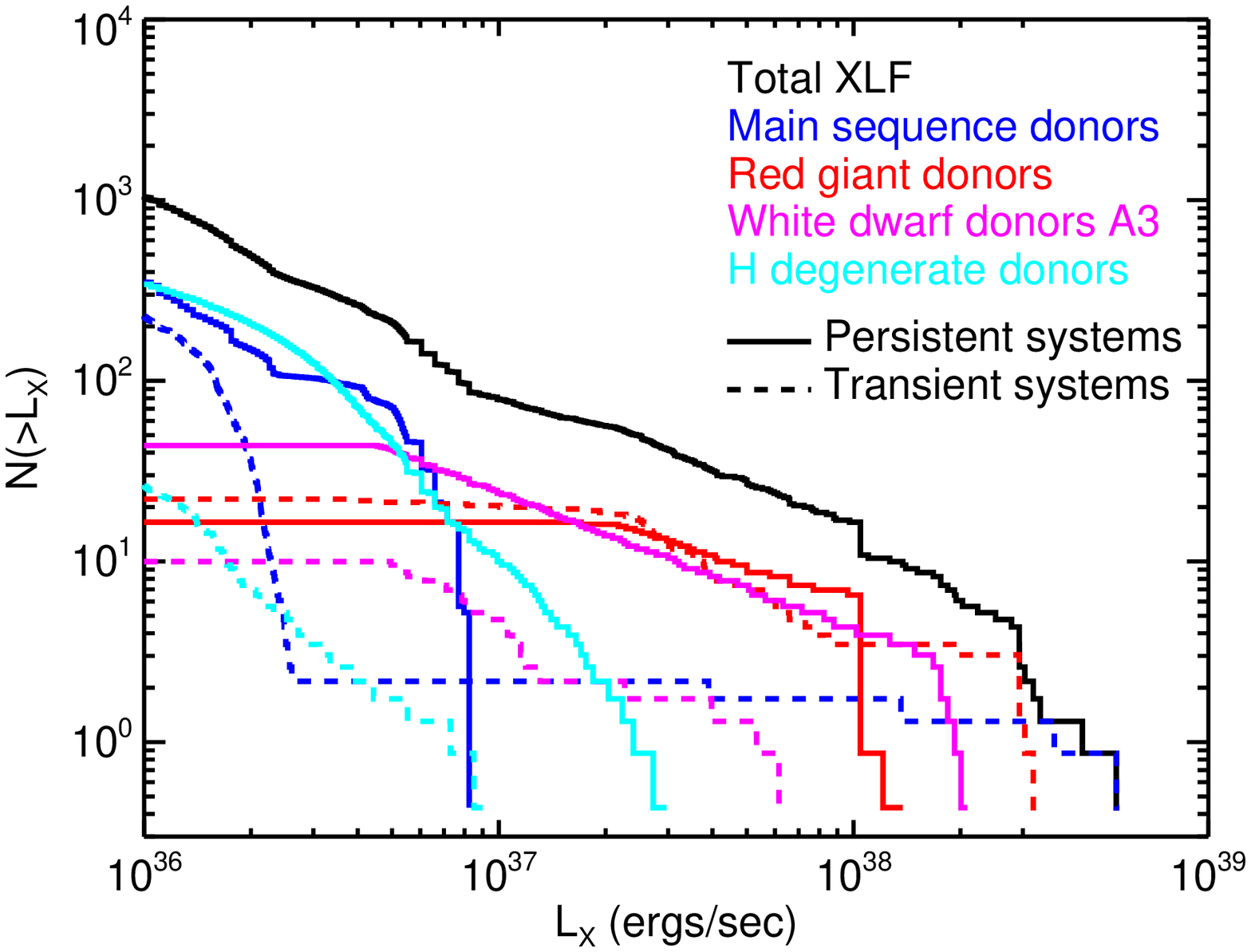}{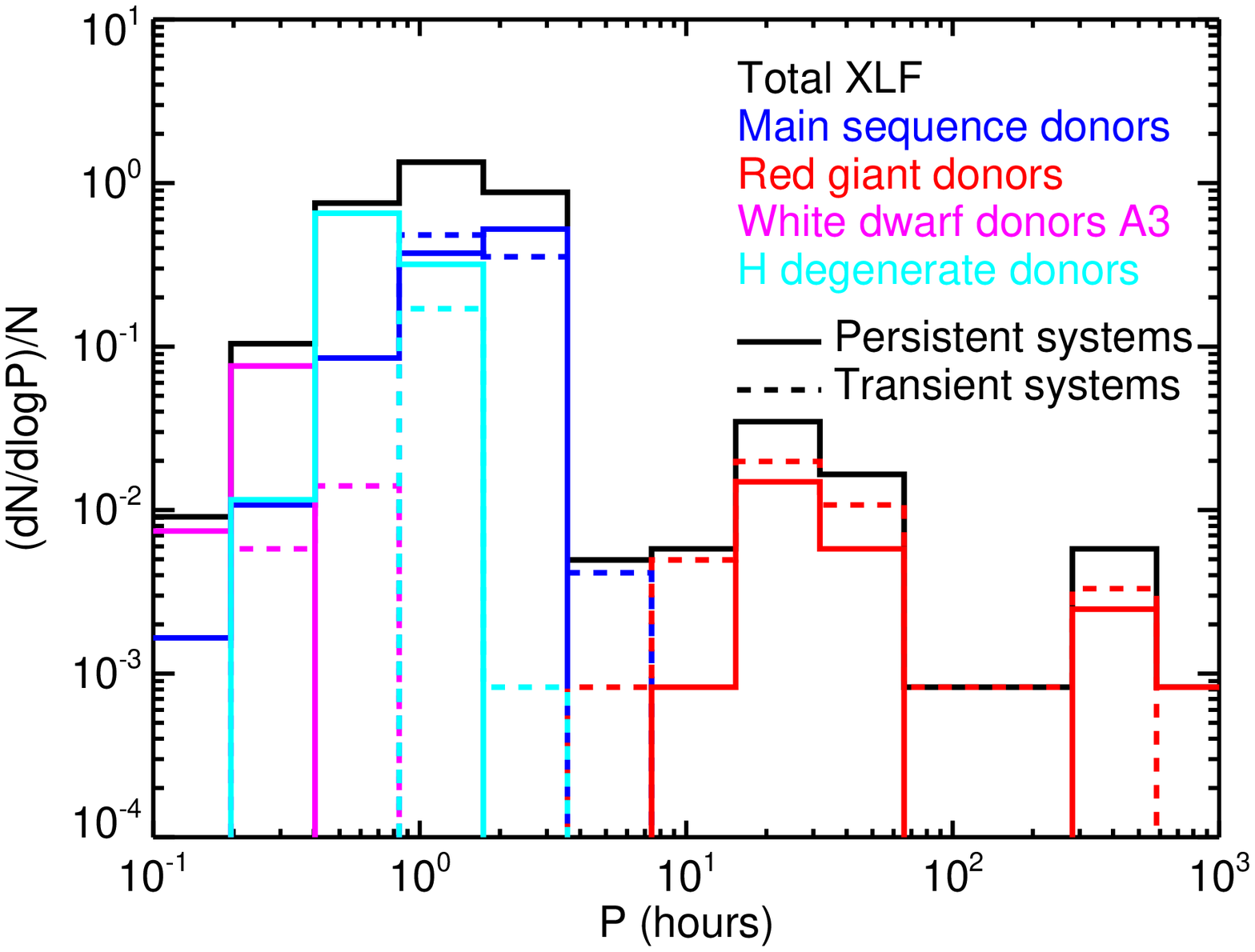}

\plottwo{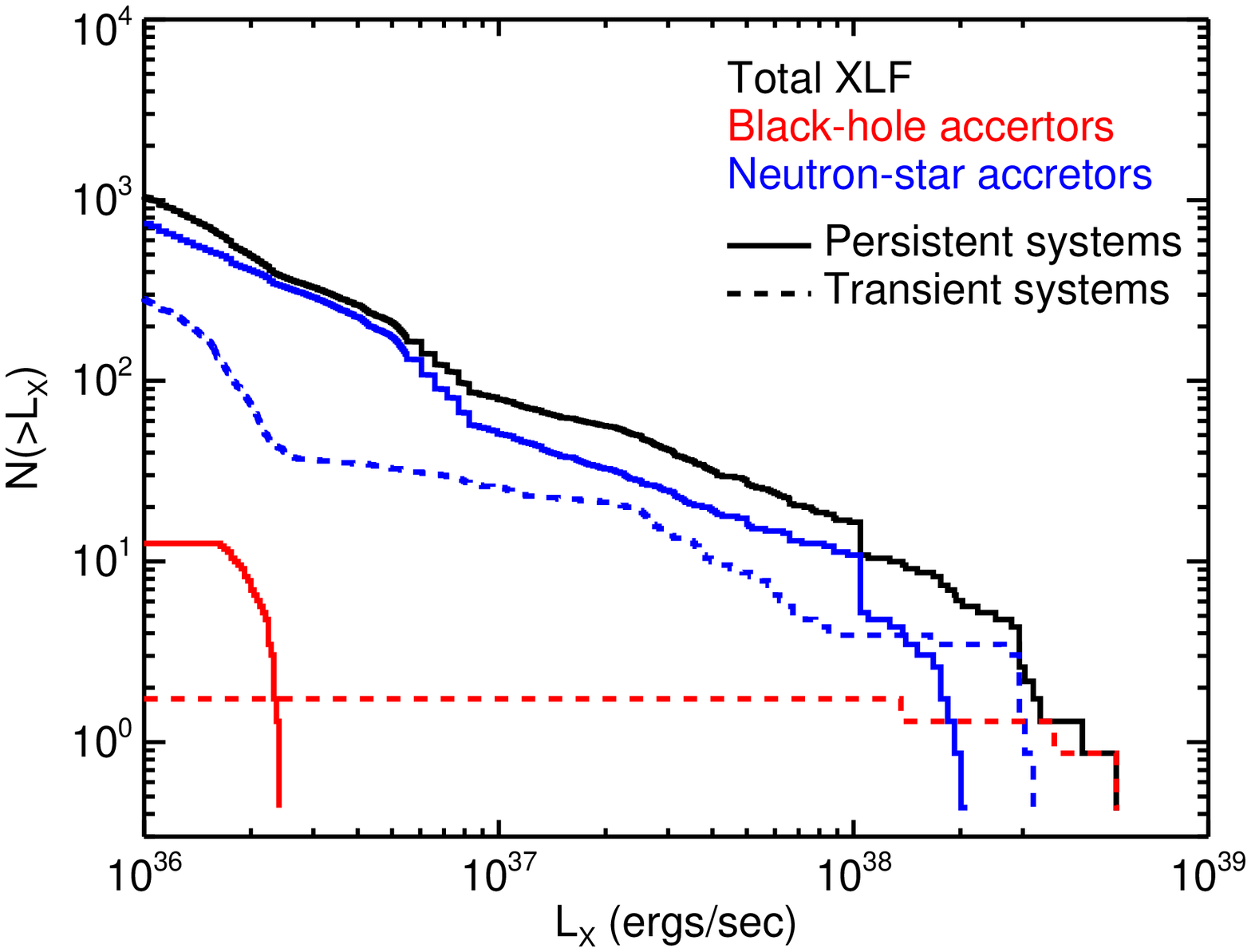}{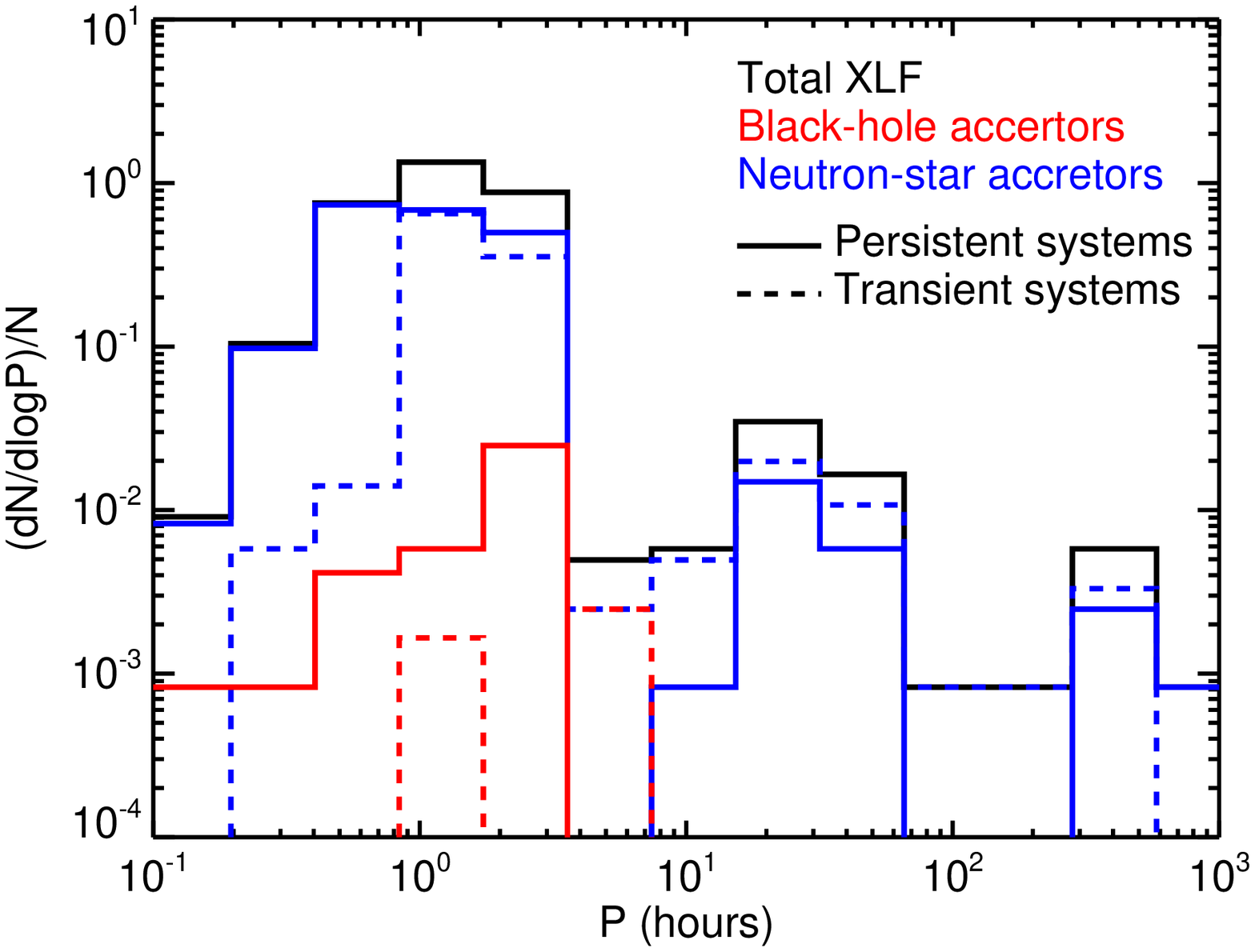}
\caption{Analysis of the LMXB population. We show the contribution of different sub-populations by separating the LMXBs into groups of systems with different donor (upper left panel) and accretor (lower left panel) stellar types. We find find that the mid-range of the XLF is dominated by NS-LMXB with red giant donors, while the high-end by BH LXBs with MS donors. Using the same separation into sub-populations we show the probability density function of the orbital period P of groups with different donor (upper right panel) and accretor (lower right panel) stellar types.}
\label{xlf_type}
\end{figure}

In order to investigate further the characteristics of our modeled LMXB population we classify the LMXBs based on their donor stellar type and also separate them in transient and persistent systems (Figure \ref{xlf_type}). We also examine the orbital periods of each sub-population. This way we can infer to what degree each sub-population contributes to the total XLF. For our \emph{standard} model $\rm 14A^{IT}$ we find that systems with MS donors are the most numerous group, but their luminosity usually falls below the observational limit ($\sim 3\times 10^{36}\rm \,erg\,s^{-1}$). The XLF is dominated by transient and persistent systems with red giant donors, which despite being less numerous overall, they are more luminous. It is worth pointing out that the population of ultra-compact LMXBs with WD donors argued by \citet{BD2004} to dominate the LMXBs in  GCs, is also present in our models. Their semi-analytically derived XLF has, as expected, the same shape as the one found in our models (power-law with a slope of -0.8). We find in our models that the number of these systems formed in the galactic fields can be comparable to the number of LMXBs with red giant donors, but most of the time their contribution to the XLF is masked by the red giant donor systems. 

LMXBs with NS accretors outnumber those with a BH accretor by a factor of $\sim 50$. A small population of transient BH-LMXB has a significant contribution only to the high luminosity end of the XLF, above $2\times 10^{38}\, \rm \,erg\,s^{-1}$. These transient BH-LMXBs have orbital periods from 1 to 10 hours and luminosity comparable to their Eddington luminosity according to eq.(\ref{Lx_mix}) we adopted in this model. We also find a sub-population of persistent BH-LMXBs with MS donors, but their luminosity falls below the observational limit of the \emph{Chandra} observations.

We performed the same analysis for all of our models (plots similar to Figure \ref{xlf_type} for all models can be found in the online supplemental material), and we identified some general trends on how the different PS parameters affect the LMXB population. The strongest effect comes from variation of the CE efficiency $\alpha_{\rm CE}$. Low CE efficiencies ($\alpha_{\rm CE}<0.3$) result to the merger of all low period systems and only the widest binaries avoid a merger and survive the CE phase. Such wide binaries become wide orbit LMXBs with red giant donors. In models with higher CE efficiencies, binaries with progressively tighter orbits survive the CE phase, and we see LMXBs with shorter period and WD or MS donors appearing in our current population. Although these systems become more numerous for $\alpha_{\rm CE}>0.4$, LMXBs with red giant donors still have the most important contribution to the XLF for luminosities above $10^{37}\, \rm \,erg\,s^{-1}$.  

The IMF affects the global picture of the LMXB population too. A flatter IMF (Salpeter) favors the formation of more massive stars. On the one hand this leads to the formation of BH-LMXBs and on the other hand favors the formation of generally more massive secondary stars which have evolved into WD by the time the system reaches the LMXB phase. We find that models with steep IMF (Scalo/Kroupa) have less LMXBs with WD donors than models with the same initial parameters but a flatter IMF (Salpeter). The stellar wind strength has similar effects on the LMXB population, as it affects the mass of the two components and thus their evolutionary state when they reach the LMXB phase. This effect is more prominent on the massive primary stars, where strong wind mass loss leads to less massive pre-supernova cores and thus fewer BH.  

  Two of the PS parameters we varied in study (IMF and stellar wind strength) greatly affect the formation of BH-LMXB. Therefore some of our models (the ones with steep IMF and strong stellar winds) turn out to be highly inefficient in the formation of BH-LMXBs.  However, a population of LMXBS with only NS accretors cannot form sources with luminosities reaching up to $10^{39}\, \rm \,erg\,s^{-1}$, unless we assume that highly super-Eddington accretion onto NS is possible. Consequently forming enough BH-LMXBs to  populate the very high end of the XLF, is a  discriminative criterion for our models.

\subsection{Time evolution of the LMXB population}
\begin{figure}
\plottwo{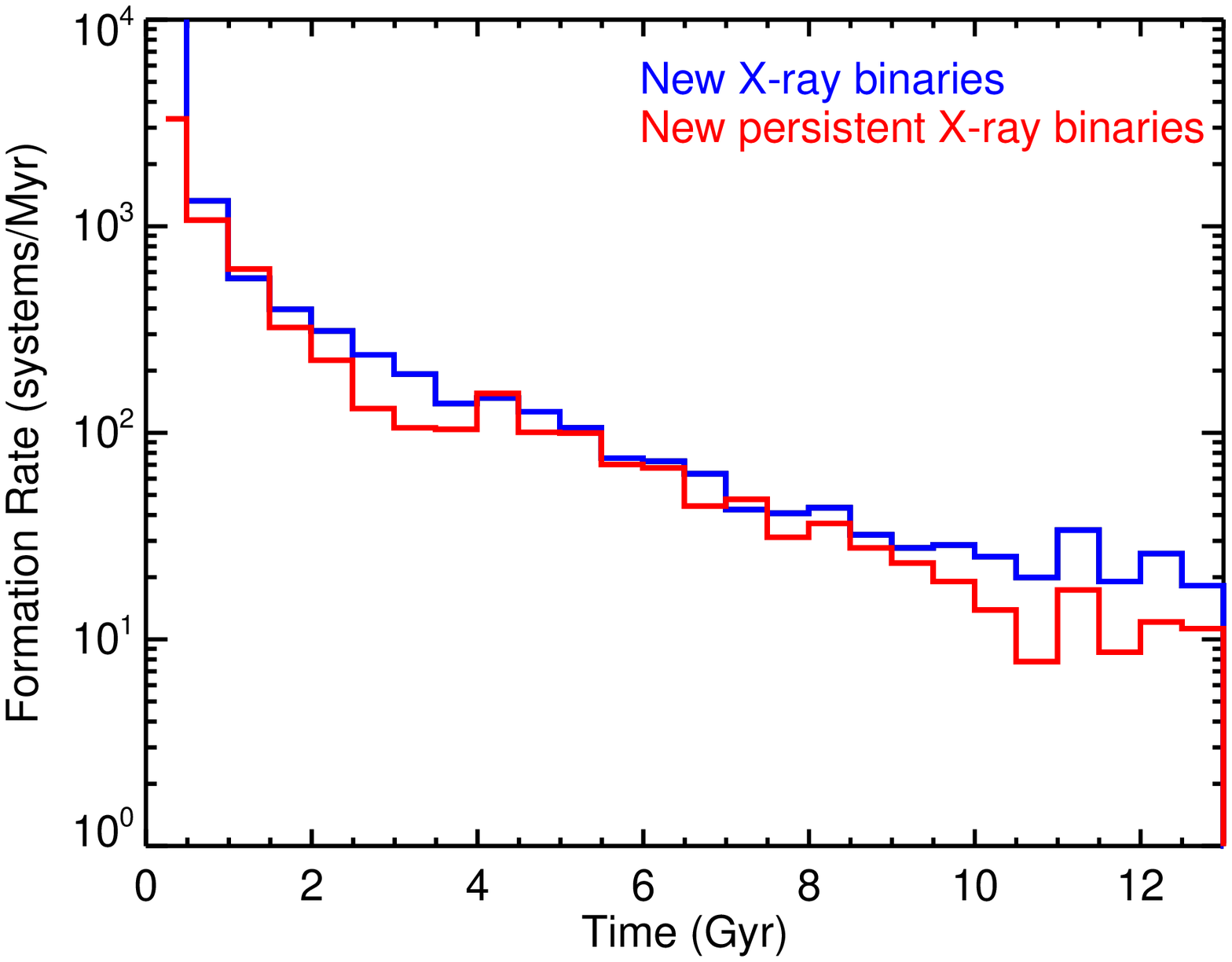}{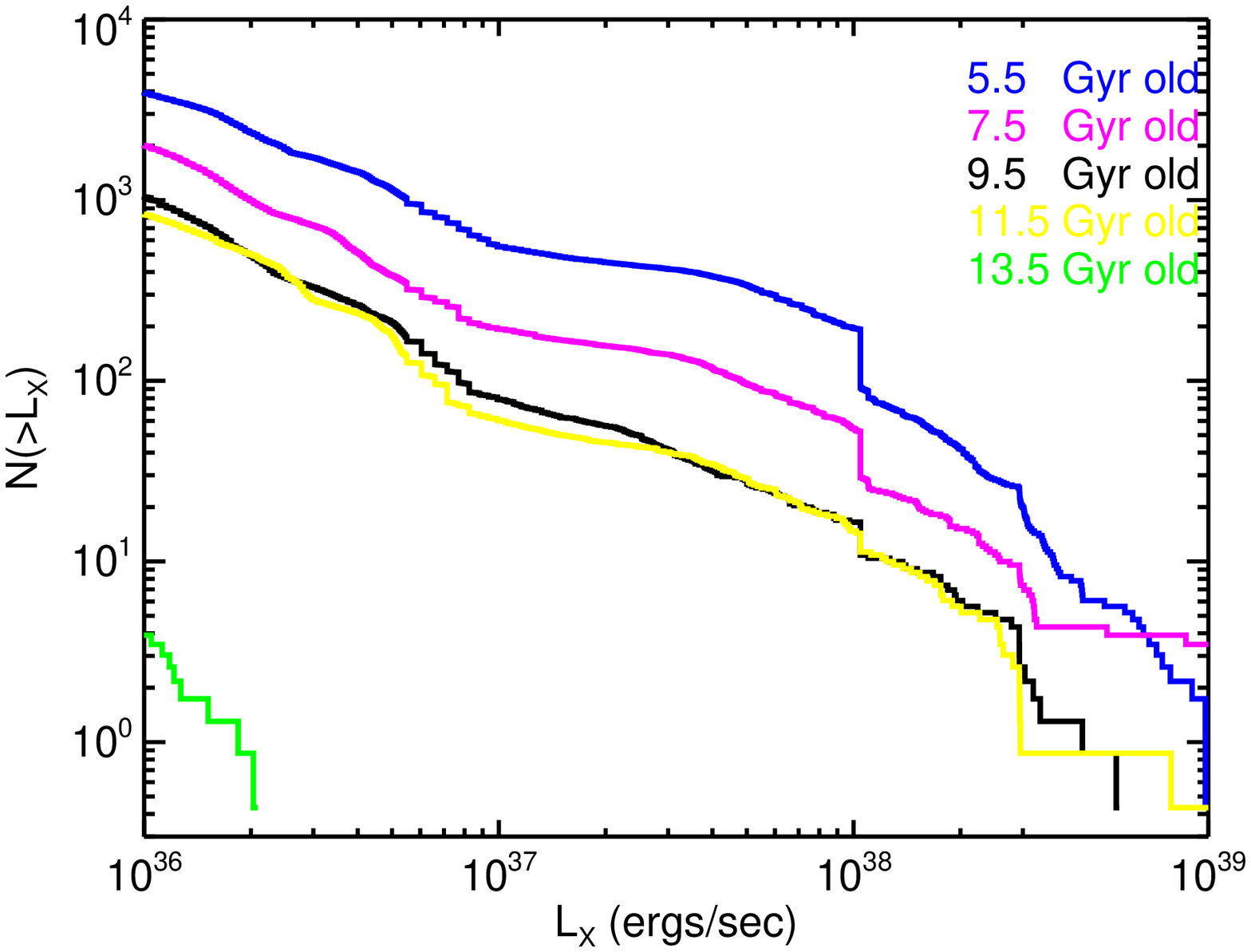}
\caption{\textbf{Left panel}: Formations rate of all x-ray binaries (blue line) and only of persistent x-ray binaries (red line)  as a function of time. After a peak in the formation rate for the first few gigayears, there is a decreasing but steady production of new persistent systems even after 10 or 11 gigayears. \textbf{Right panel}: Evolution of the XLF with the age of the galaxy.}
\label{xlf_age}
\end{figure}

The age of the population is a very important factor in our simulations, as the characteristics of the binaries and their relative numbers and contributions vary significantly over time.  In the right panel of Figure \ref{xlf_age} we see the evolution of the modeled XLF versus time. The total number of sources as well as the number of luminous sources decreases steadily with time. At early times, the first 5 to 6\,Gyr, the XLF has significant contributions from intermediate mass X-ray binaries. The increased number of luminous sources at earlier times makes the shape of the XLF flatter. On the other extreme, in a 13 to 14\,Gyr old stellar population, a population of luminous LMXBs is almost non-existent. The formation rate of X-ray binaries in general and of persistent sources in particular can be seen in right panel of Figure \ref{xlf_age}. The two curves follow closely each other, which means that according to our models most the X-ray binaries go through a phase of persistent emission at some point of their evolution. Another characteristic that initially might seem counter-intuitive is that after a peak in the production of X-ray binaries in the first few gigayears, there is a decreasing but \emph{steady} production of \emph{new} persistent systems even after 10 or 11\,Gyr. This suggest that there can be a very long delay between the formation of the compact object (at $\sim 1-100 \rm \,Myr$) and the onset of the mass-transfer phase. We should note that the derivation of the ages of the two galaxies by \citet{TF2002} involves a systematic uncertainty of at least 1\,Gyr. This uncertainty in the age is an additional factor that can account for the difference in the number of observed sources between the two galaxies and of course some deviation between the observed and the modeled XLFs.

\subsection{The implication of GCs in the LMXB population: NGC3379 vs. NGC4278.}

The simulations we present  in this initial study do not take into account formation channels involving dynamical interactions, believed to dominate the formation of LMXBs in GCs. However, we know from comparisons of optical and X-ray observations that 10-70\% of LMXBs in early type galaxies are located in GCs, depending on the type of the galaxy and the GC-specific frequency \citep{KMZ2007, Jordan2004, Sarazin2003}. It is thus clear that at least for some galaxies, the LMXB formation in  GC has an important contribution to the total population. \citet{Fabbiano2007} studied the GC LMXB population of NGC3379 and found that only 15\% of all LMXBs detected in this galaxy is coincident with GC. Furthermore, they argued that there are differences in the XLF of the two populations of field and GC LMXBs, but \emph{only} at low luminosities (below $10^{37}\rm \, erg\, s^{-1}$) where the incompleteness of the observed population does not allow reliable direct comparison with theoretical models. Thus, it turns out that it is valid to compare our models to the total observed XLF (without subtracting the GC LMXBs) of NGC337, for luminosities above $10^{37}\rm \, erg\, s^{-1}$. The case of NGC4278 is different as its GC-specific frequency is 5 times larger than that of NGC3379, and the contribution of GC LMXBs is expected to be more significant there. An analysis similar to that of \citet{Fabbiano2007} for NGC3379, will appear  in the near future and it will enable us to separate the field from the GC LMXB population in our comparison. 

 The realistic modeling of X-ray binaries formed in GC requires the coupling of cluster dynamics with binary evolution. \citet{BD2004} suggested that ultra-compact X-ray binaries dominate the cluster population and they are continuously replenished through dynamical interactions. These systems initially evolve through a bright persistent phase (for $3\,\rm Myr$), then their mass transfer rate drops gradually as the WD donor mass decreases; the systems become transients and their contribution to the XLF is diminished.  Our models for formation of LMXBs from primordial binaries produce a population of ultra-compact binaries too. Not surprisingly, since the properties of the binaries are very similar, the XLF of this sub-population has the exact same shape as the one produced by the model of \citet{BD2004}:  a power law shaped XLF (slope of -0.8), slightly flatter than the observed XLF of NGC3379 and NGC4278. In the case of the galactic field population though, there is no enhancement in the formation rate of these systems via dynamical interactions. So our models do not produce enough luminous persistent systems to dominate the total XLF. Their contribution, although significant, it is usually masked by LMXBs with red giant donors, which tend to have higher X-ray luminosities.  More recently, \citet{Ivanova2007} studied the formation of compact binaries containing NSs in dense GCs. In their simulations they used \emph{StarTrack} as their PS model and adopted a simplified treatment of the dynamical events \citep[see also][]{Ivanova2006}. They found that formation rate of dynamically formed LMXBs with red giant and MS donors is comparable to the ultra-compact X-ray binaries formed in GC, but the later ones have significantly shorter lifetimes. This last result casts doubts on the suggestion by \citet{BD2004} for an ultra-compact X-ray binary dominated population in GCs.

\section{Conclusions and Discussion}

The recent deep \textit{Chandra} observations \citep{Kim2006} of the two typical old elliptical galaxies: NGC3379 and NGC4278, led to the first observed low-luminosity XLFs of LMXBs, with the detection limit ($3\times 10^{36}\rm \, erg\, s^{-1}$) being about an order of magnitude lower than in most previous surveys. Motivated by this observational work, we developed PS simulations of LMXBs appropriate for these two galaxies. We considered formation of LMXBs only through the evolution of primordial binaries in the galactic fields and examined the possible contribution to the overall LMXB population. For our modeling we used the updated PS code \emph{StarTrack} \citep{Belczynski2006}.

Our main conclusions can be summarized as follows:

We found that some of our models produce XLFs in very good agreement with the observations, based on both the XLF shape and absolute normalization. There is no unique combination of PS parameters and modeling of transient sources (DC and outburst luminosity) that gives an XLF in agreement with the observation. We conclude that formation of LMXBs in the galactic field via evolution of primordial binaries can have a significant contribution to the total population of an elliptical galaxy, especially the ones with low GC specific frequency such as NGC3379 \citep{Fabbiano2007}. Nevertheless, we are able to exclude the majority of our models, as inconsistent with the observations. Note that widely used, simple assumptions such as that all transients source in the outburst state are emitting X-rays at $L_{\rm Edd}$, lead to XLFs clearly inconsistent with the observed ones. Our results appear to be robust since we do not have to fine-tune our code parameters in order to get a model that resembles the observed population.

As already suggested by \citet{Piro2002}, the LMXB population has a significant contribution by transient systems (thermal disk instability) and with reasonable outburst  DCs they can even dominate the XLF. As a consequence the XLF shape is rather sensitive to the treatment of these transient systems. In Figure \ref{xlf_dc} we show that keeping the same PS parameters and changing \emph{only} the modeling of transient sources leads to completely different XLFs. We tried different methods of modeling the outburst characteristics of transient LMXBs and we found that: \emph{(i)} When we assume the outburst luminosity of all transient LMXBs to be equal to $L_{\rm Edd}$ (transient treatment E) or apply eq.(\ref{Lx_P}) (transient treatment F) - which was empirically derived for Galactic BH-LMXBs - to the whole population, we get XLFs inconsistent with the observed ones regarding both their shape and the total number of sources predicted. \emph{(ii)} A constant DC for all systems, although not physically motivated, can be sometimes a good first approximation. \emph{(iii)} We get the best agreement with observations, when we consider a variable DC for NS-LMXB based on the theoretical study of \citet{DLM2006} (see eqs. \ref{dc} and \ref{Lx_phys}), while for the BH-LMXBs we use the empirical correlation between orbital period and outburst luminosity, derived by \citet{PDM2005} (see eq. \ref{Lx_P}), and assuming a low constant DC ($\sim 5\%$).     

The LMXB sub-populations that mainly contribute to the model XLFs are NS-LMXBs with red giant donors and BH-LMXBs with MS donors. A population of persistent ultra compact LMXBs with WD donors is also present in our models and in some cases has an important contribution too (see models $\rm 9A^{IT}$, $\rm 11A^{IT}$ and $\rm 12A^{IT}$).  Of these sub-populations, the NS-LMXBs are the most dominant and primarily determine the XLF shape in the medium and low luminosity range (below $2 \times 10^{38}\rm \, erg \, s^{-1}$), while the BH-LMXBs have a significant contribution to the high-end of the XLF.

The normalization of the modeled XLFs is a less robust characteristic than its shape. We normalize the models so that the number of the primordial binaries we evolve, correspond to the known galaxy masses, given the IMF and the binary fraction. There are however uncertainties of the order of a few in the determination of the mass of the observed galaxies, due to uncertainties in their distance, the bolometric luminosity and the light-to-mass ratio. The majority of the models presented here  produce the observed number of LMXBs to within a factor of 3, consistent with the galaxy mass uncertainties. Exceptions are  models with transient treatment E or F (see Table \ref{models_tran}) and models with high CE efficiencies (models 21-28, see \ref{models}) which greatly overproduce hight luminosity LMXBs. Furthermore small changes in the CE efficiency can change the total number of sources produced by a model by a factor of two, without changing significantly the shape of the XLF (compare for example models $\rm 10A^{IT}$, $\rm 14A^{IT}$, $\rm 18A^{IT}$ in the online supplemental material). In view of these uncertainties and our limited parameter space exploration for the models, we consider this normalization agreement satisfactory, but do not use it as an actual constraint on the models. 

We do not claim that the work presented in this Paper is a complete PS study of field LMXBs in elliptical galaxies. It is meant to be a first effort in interpreting the recent deep \emph{Chandra} observations of the two elliptical galaxies NGC3379 and NGC4278. Throughout the Paper we identify the caveats of our analysis, and we intend to address them in our future work. It turns out that the realistic treatment of the outburst properties of transient LMXBs is crucial for the modeling of XLFs of extragalactic populations. The derivation of an empirical correlation between the outburst luminosity and the period of Galactic NS-LMXBs, similar to the one derived by \citet{PDM2005} for Galactic BH-LMXBs,  will provide a better understanding of the differences in the transient behavior of these two classes of X-ray sources. Another simplifying assumption we made, is the constant outburst luminosity of transient LMXBs. The use of model lightcurves  for the outburst phases of transient XLFs will possibly affect the shape of the modeled XLFs. We note that most of our models produce many systems with X-ray luminosity below the observational limit and down to $10^{35}\rm \, erg \, s^{-1}$. The integrated diffuse X-ray emission from these galaxies can put a strong constraint on our models. The total luminosity of the observed diffuse emission will also include emission from gas and stellar coronae \citep{PF1994, Revnivtsev2007} and thus  must be higher than the integrated luminosity of all the LMXBs in our models with luminosities below the observational detection limit.      

\clearpage
\newpage

\section*{Statistical analysis and XLFs for the complete set of models (to appear as online supplemental material)}

\begin{deluxetable}{ccccccccc}
\tablecolumns{1}
\tabletypesize{\scriptsize}
\tablewidth{0pt} 
\tablecaption{Statistical Comparison of Models to the Observed XLF of NGC4278
\label{stat_test4278}} 
\tablehead{ \colhead{Model} & 
			\multicolumn{4}{c}{NGC 3379} &
			\multicolumn{4}{c}{NGC 4278} \\
	 \colhead{}	&	
     \colhead{$P_{\rm KS}$} & 
     \colhead{$P_{\rm Kuiper}$} & 
     \colhead{$P_{\rm \chi^2}$} &
     \colhead{$P_{\rm RS}$} & 
     \colhead{$P_{\rm KS}$} & 
     \colhead{$P_{\rm Kuiper}$} & 
     \colhead{$P_{\rm \chi^2}$} &
     \colhead{$P_{\rm RS}$}  

}
\startdata
\mbox{$\rm 1A^{RVJ}$}    &    2.85e-11    &    6.19e-10    &    5.96e-08    &    1.33e-07    &    9.52e-09    &    1.74e-08    &    3.76e-05    &    6.50e-08    \\
\mbox{$\rm 1B^{RVJ}$}    &    7.41e-12    &    1.36e-10    &    0.00e+00    &    5.02e-08    &    2.82e-09    &    3.50e-08    &    3.10e-06    &    5.58e-08    \\
\mbox{$\rm 1C^{RVJ}$}    &    2.51e-17    &    4.80e-17    &    0.00e+00    &    2.71e-14    &    8.12e-15    &    6.96e-15    &    0.00e+00    &    1.79e-11    \\
\mbox{$\rm 1D^{RVJ}$}    &    3.00e-12    &    4.18e-12    &    0.00e+00    &    2.79e-13    &    2.73e-09    &    4.99e-10    &    0.00e+00    &    2.39e-08    \\
\mbox{$\rm 1E^{RVJ}$}    &    9.61e-22    &    7.91e-22    &    0.00e+00    &    0.00e+00    &    1.46e-20    &    1.78e-20    &    0.00e+00    &    0.00e+00    \\
\mbox{$\rm 1F^{RVJ}$}    &    3.30e-20    &    3.12e-20    &    0.00e+00    &    0.00e+00    &    1.15e-19    &    1.82e-22    &    0.00e+00    &    0.00e+00    \\
\mbox{$\rm 1A^{IT}$}    &    2.77e-10    &    7.57e-10    &    0.00e+00    &    7.93e-10    &    2.63e-04    &    1.21e-04    &    5.53e-04    &    1.34e-05    \\
\mbox{$\rm 1B^{IT}$}    &    1.59e-09    &    1.10e-08    &    0.00e+00    &    5.23e-07    &    7.33e-05    &    2.11e-05    &    3.50e-04    &    6.18e-07    \\
\enddata

\end{deluxetable}

\begin{figure}
\plotone{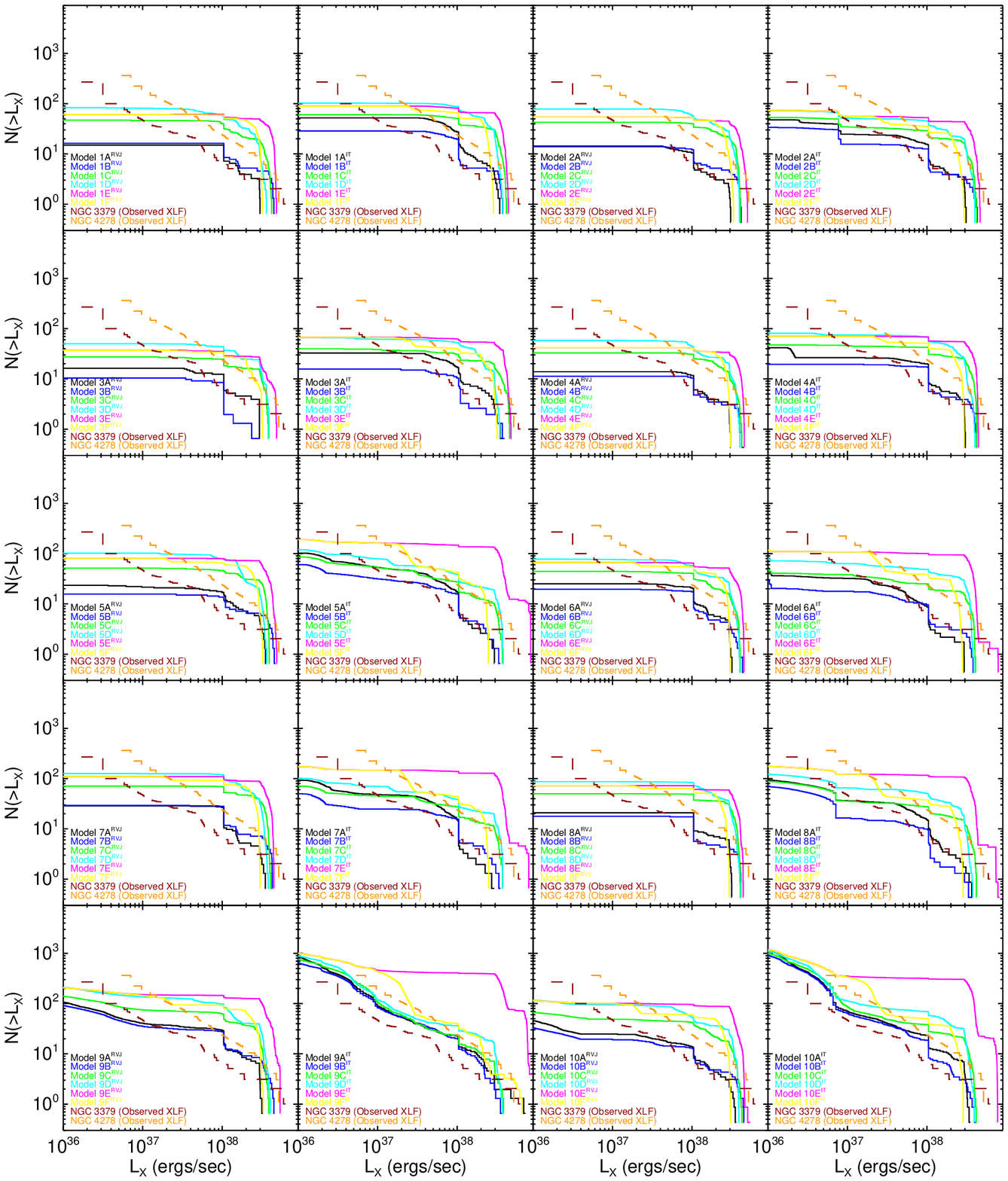}
\end{figure}

\begin{figure}
\plotone{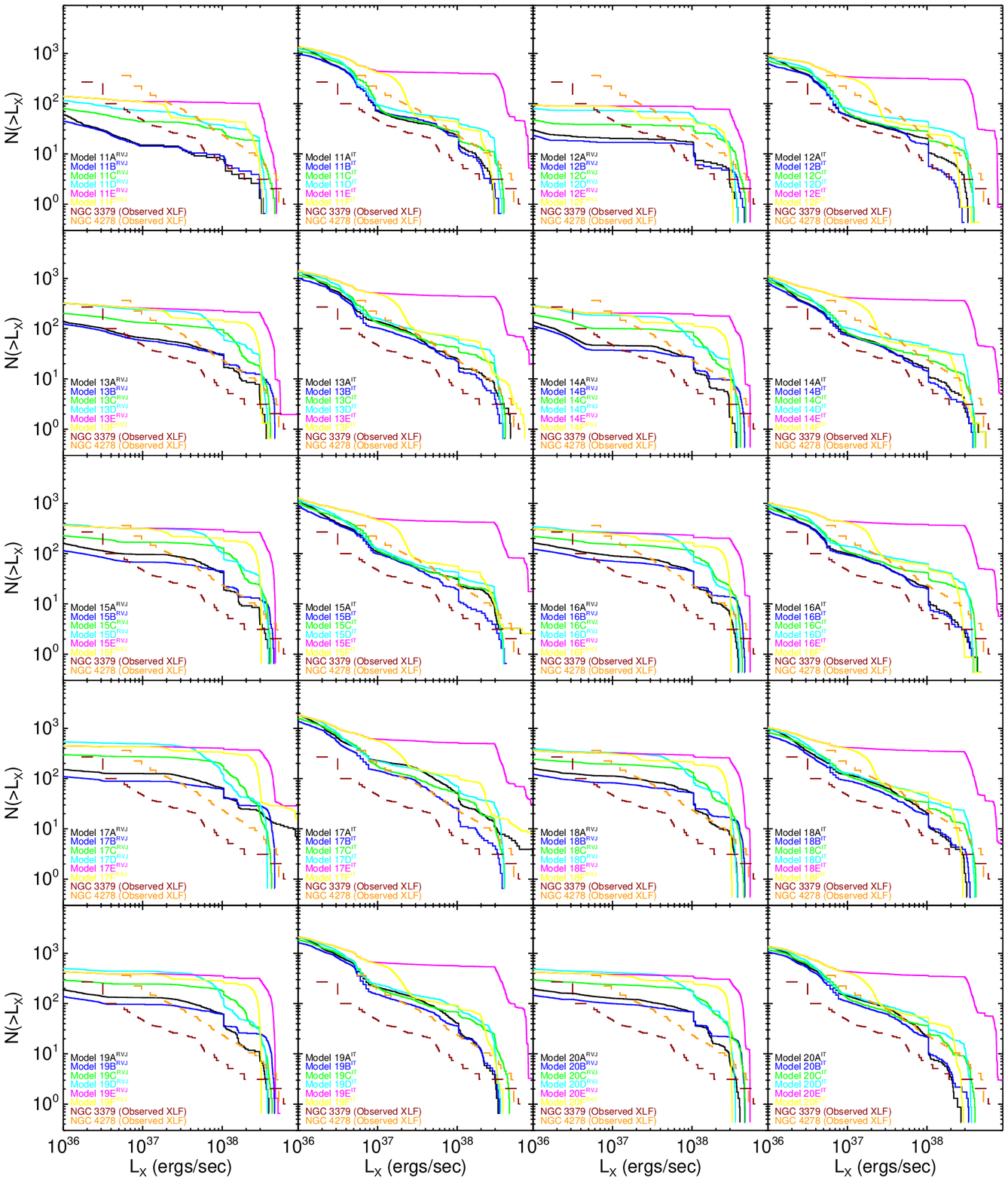}
\end{figure}

\begin{figure}
\plotone{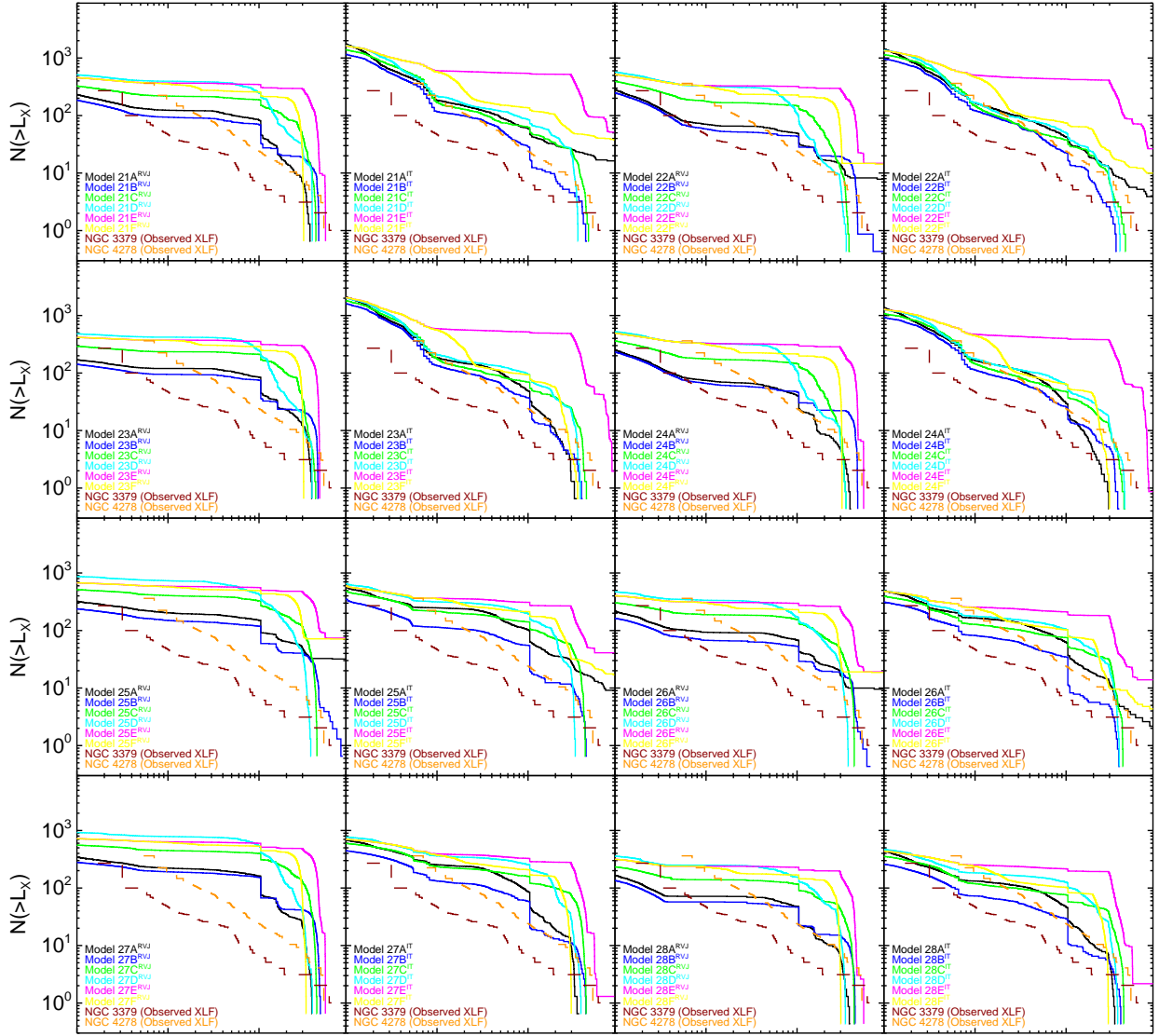}
\caption{Model XLF for all the different LMXB population models listed in Tables \ref{models} and \ref{models_tran}. In each panel all the PS parameters are kept constant, and only the modeling of transient systems changes. For comparison the observed XLFs of NGC3379 (dark red) and of NGC4278 (orange) are drawn.}
\label{xlf_dc_multi}
\end{figure}

\begin{figure}
\plotone{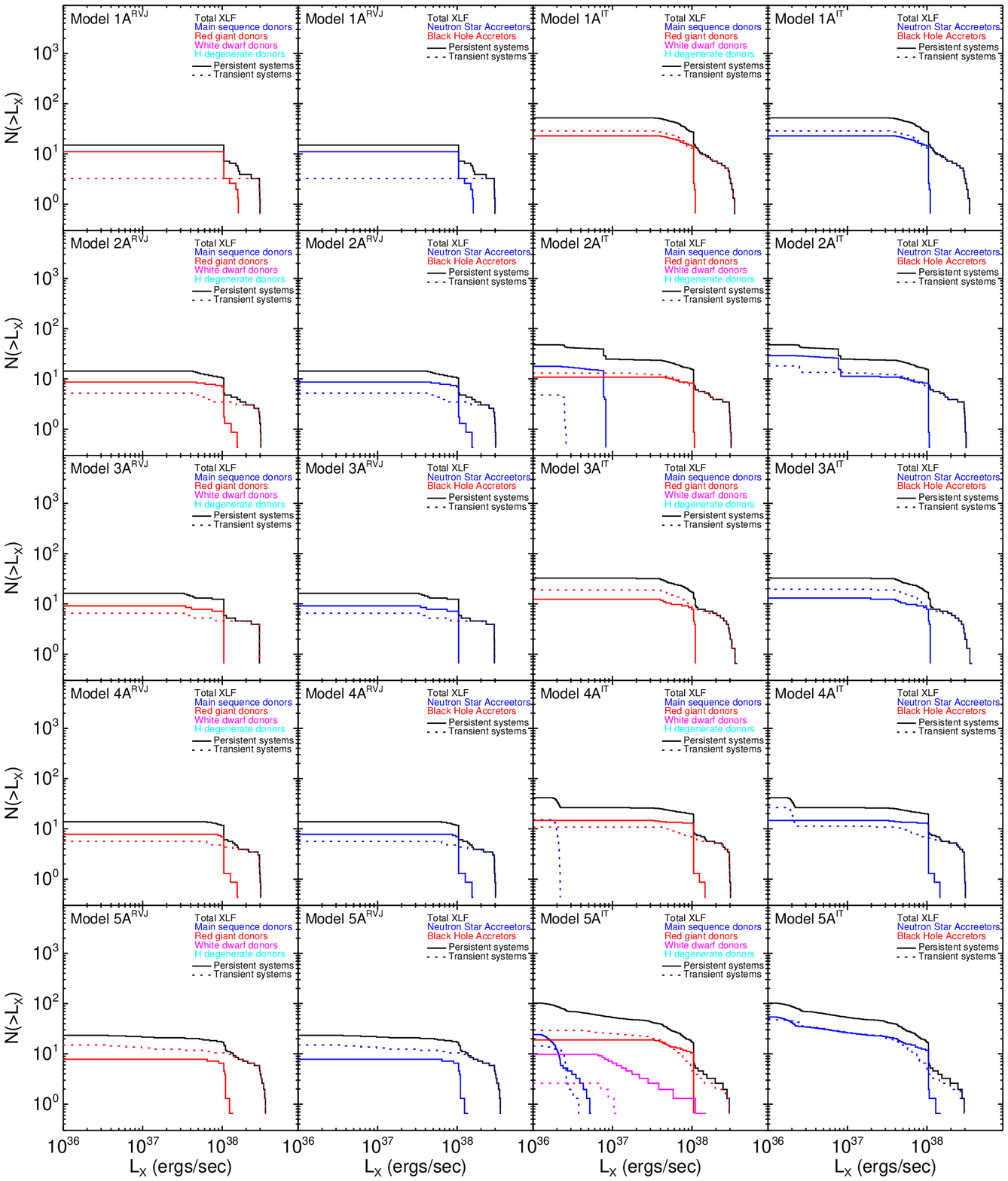}
\end{figure}

\begin{figure}
\plotone{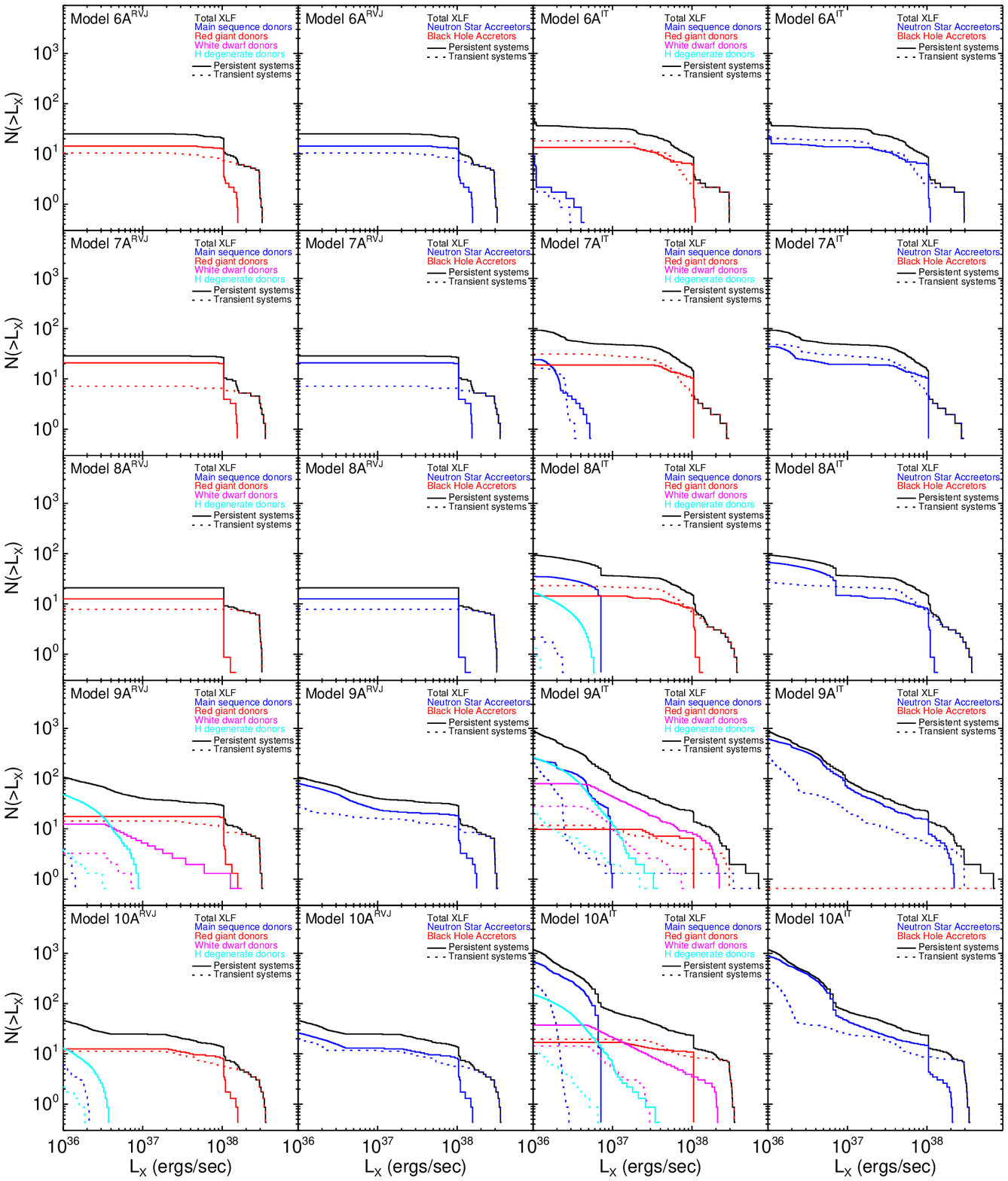}
\end{figure}

\begin{figure}
\plotone{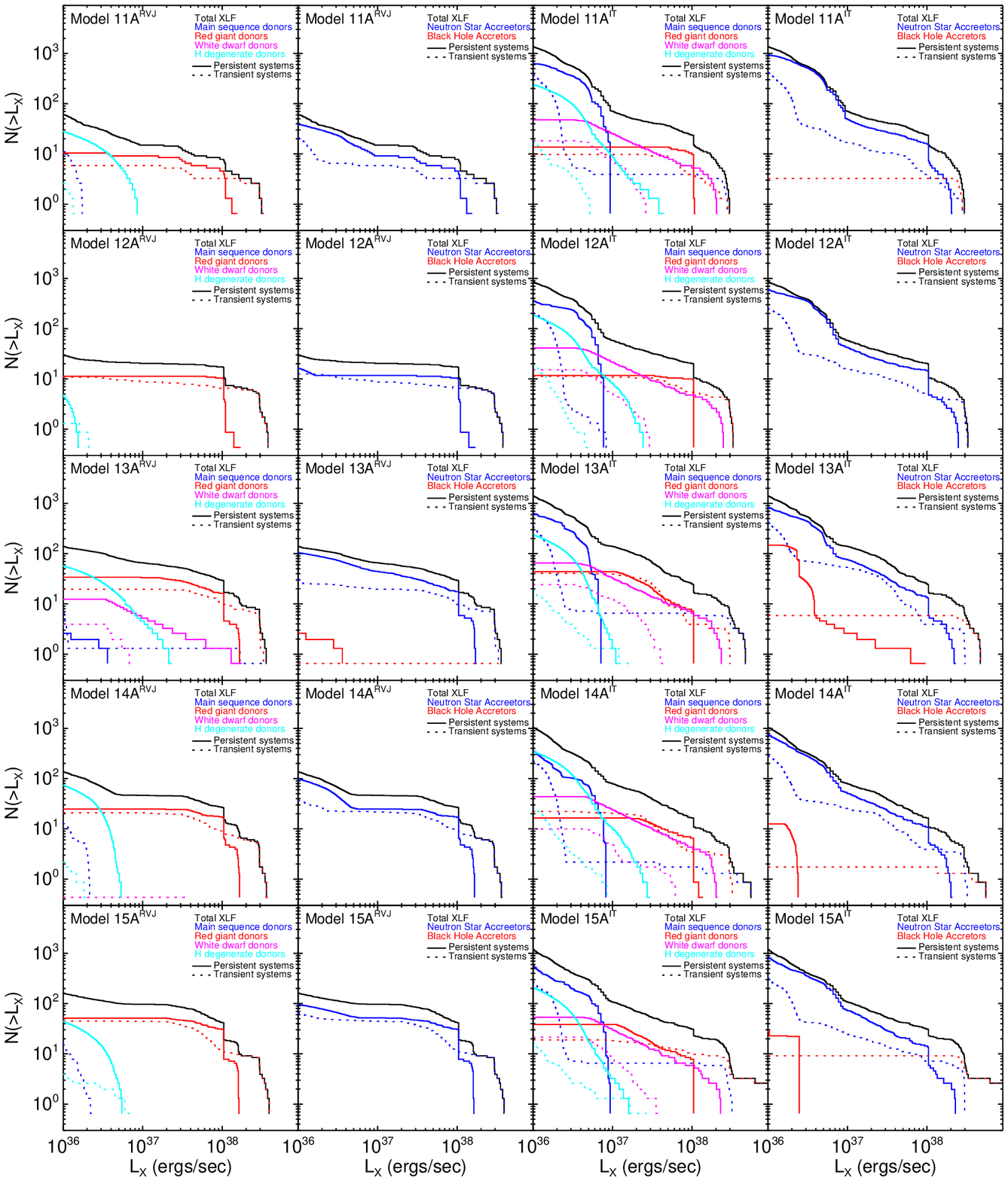}
\end{figure}

\begin{figure}
\plotone{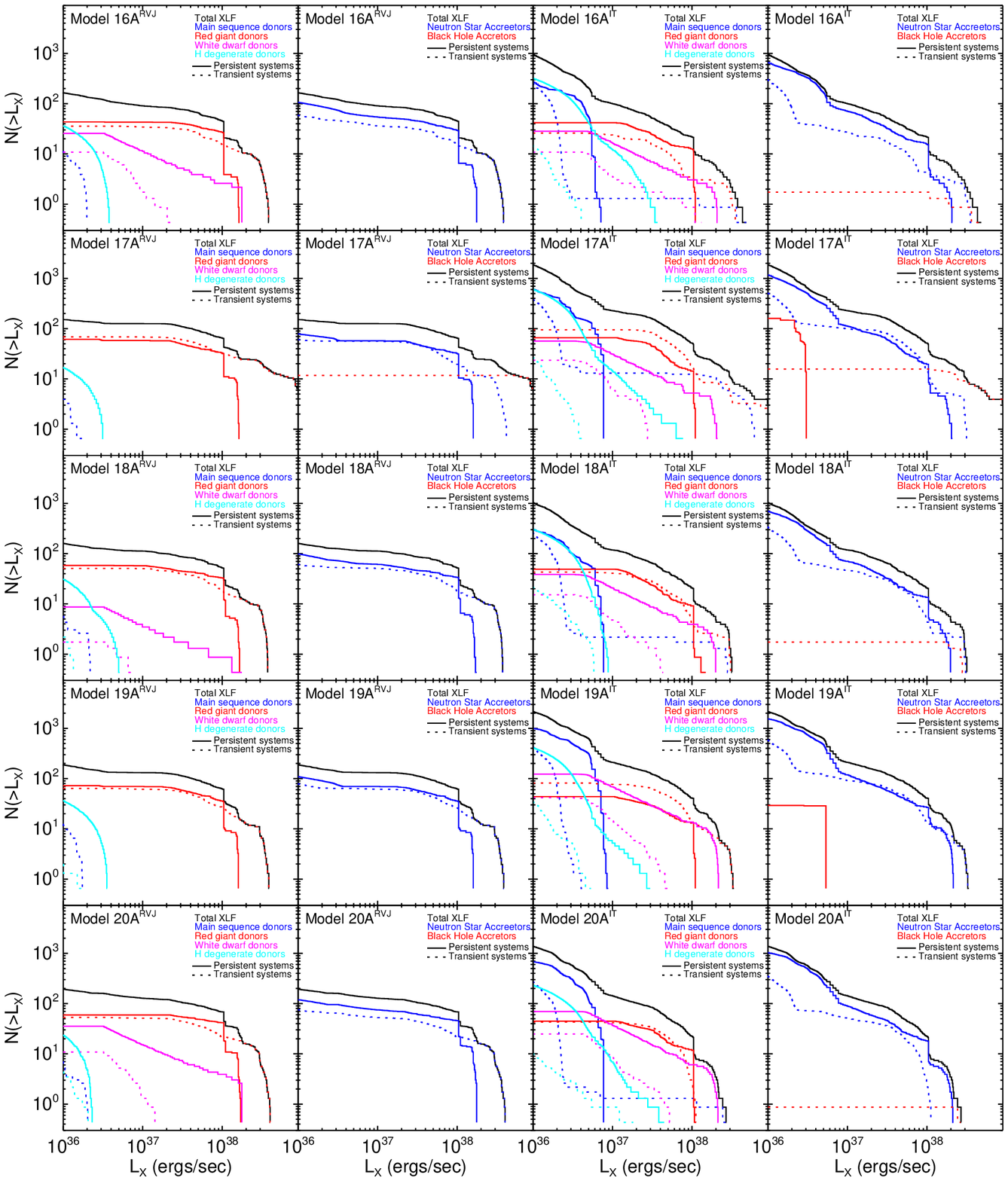}
\end{figure}

\begin{figure}
\plotone{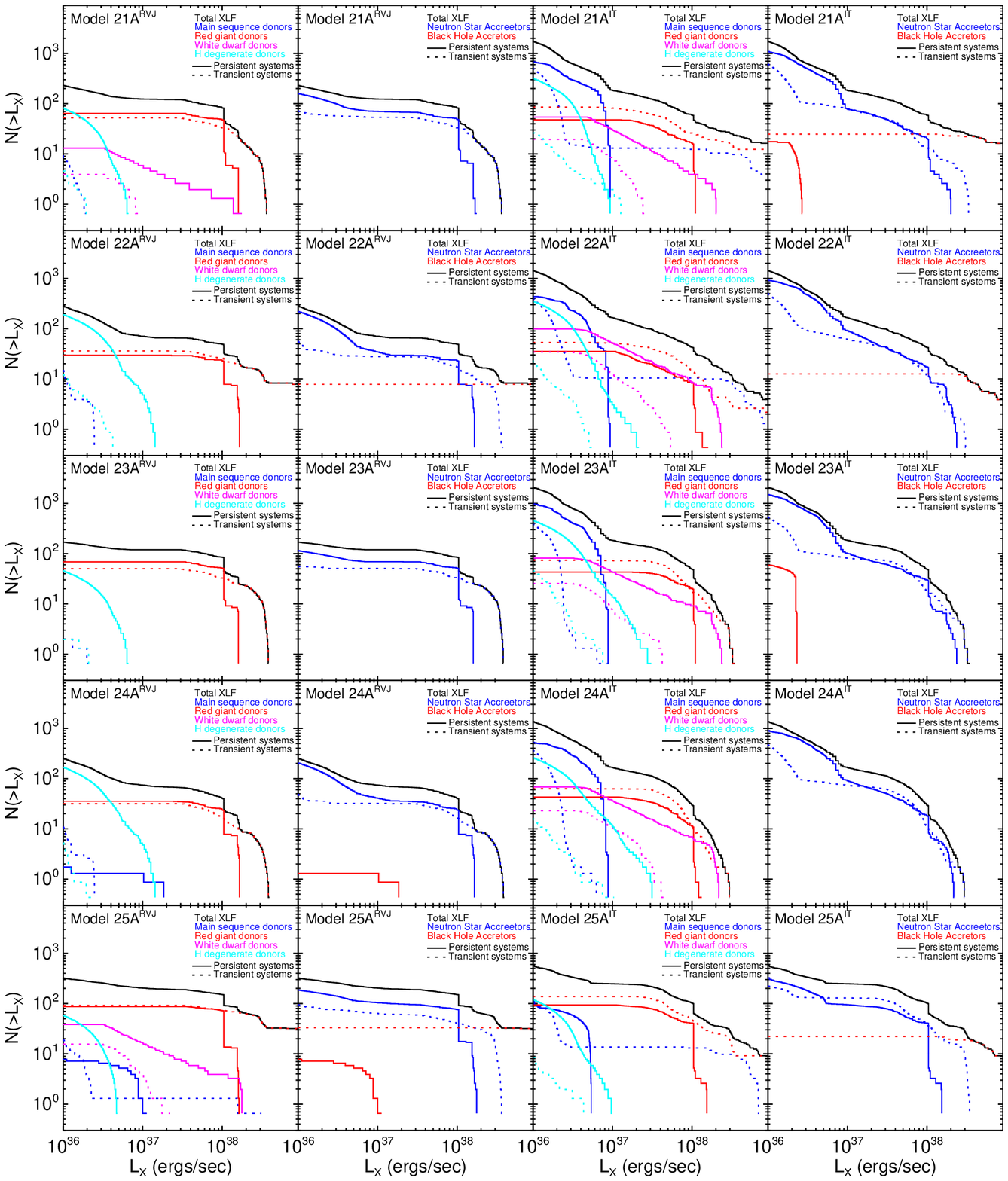}
\end{figure}

\begin{figure}
\plotone{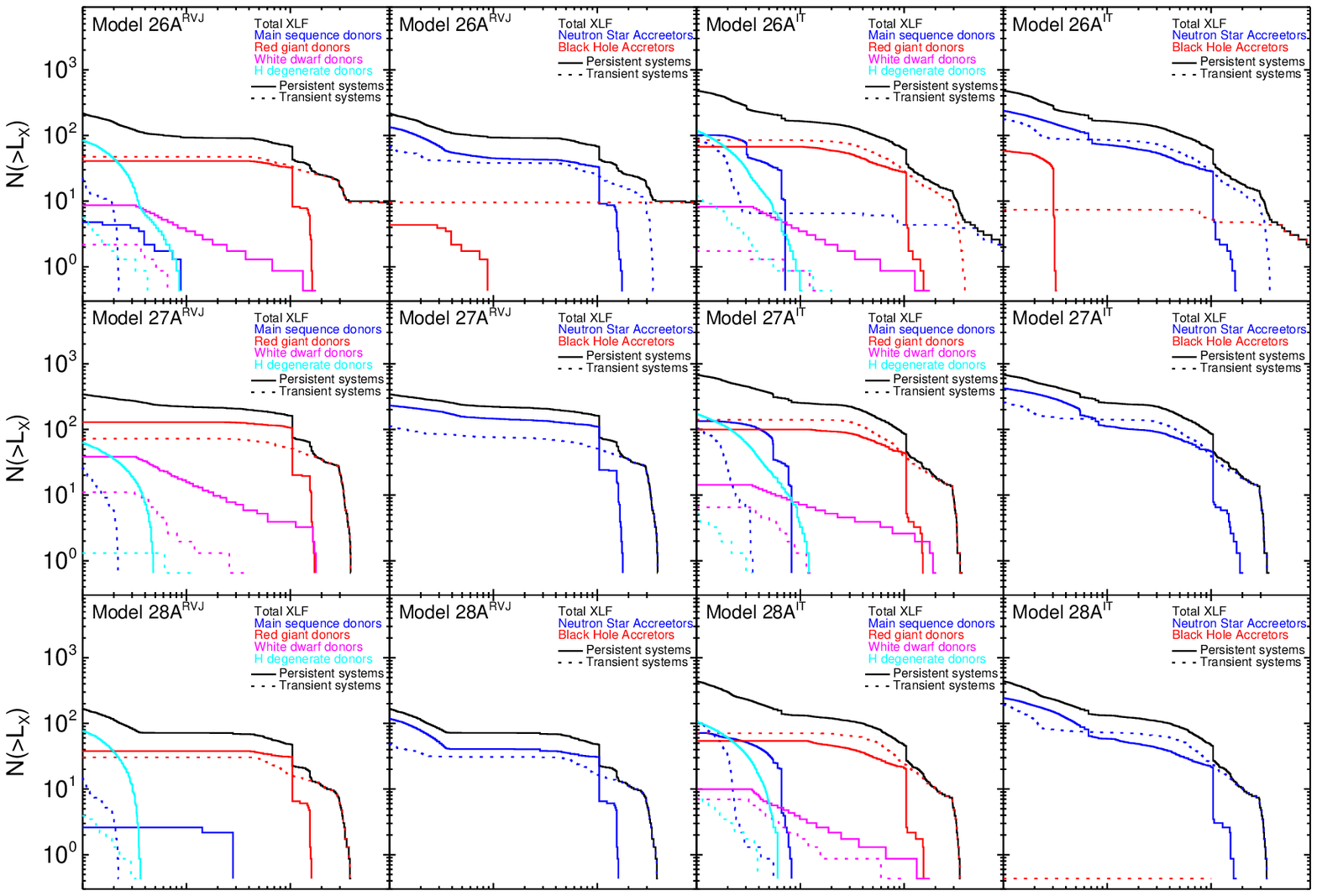}
\end{figure}

\begin{figure}
\plotone{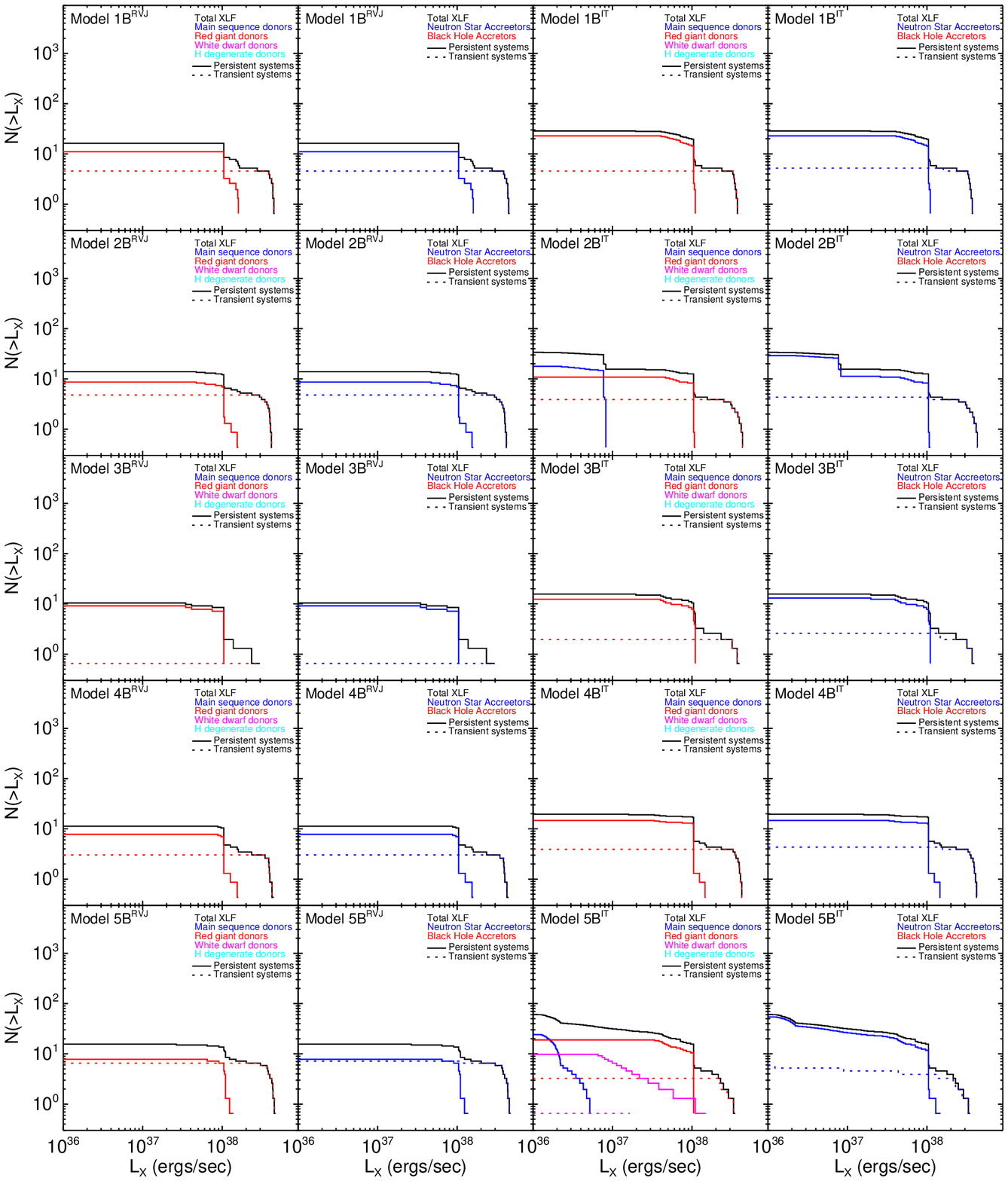}
\end{figure}

\begin{figure}
\plotone{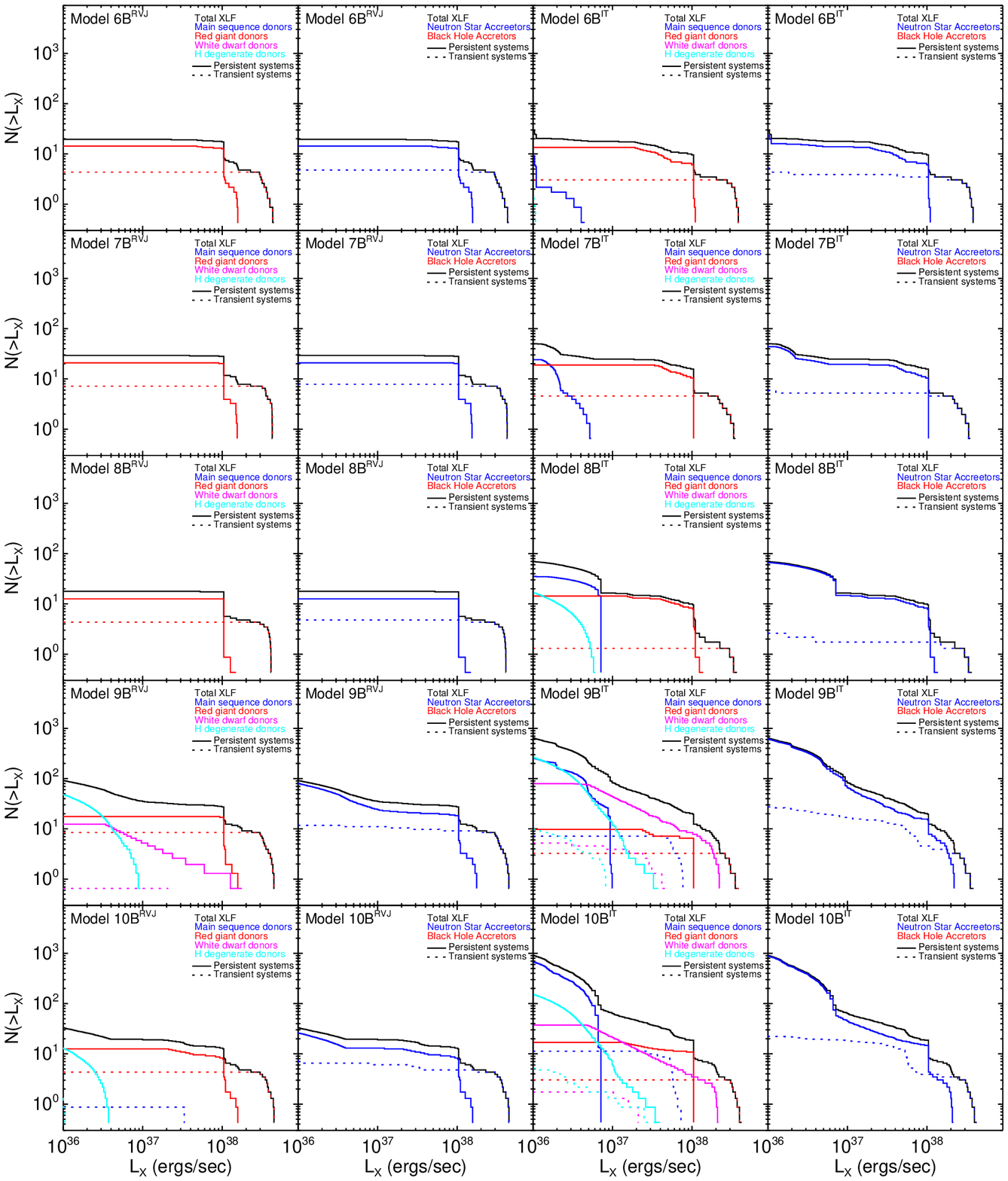}
\end{figure}

\begin{figure}
\plotone{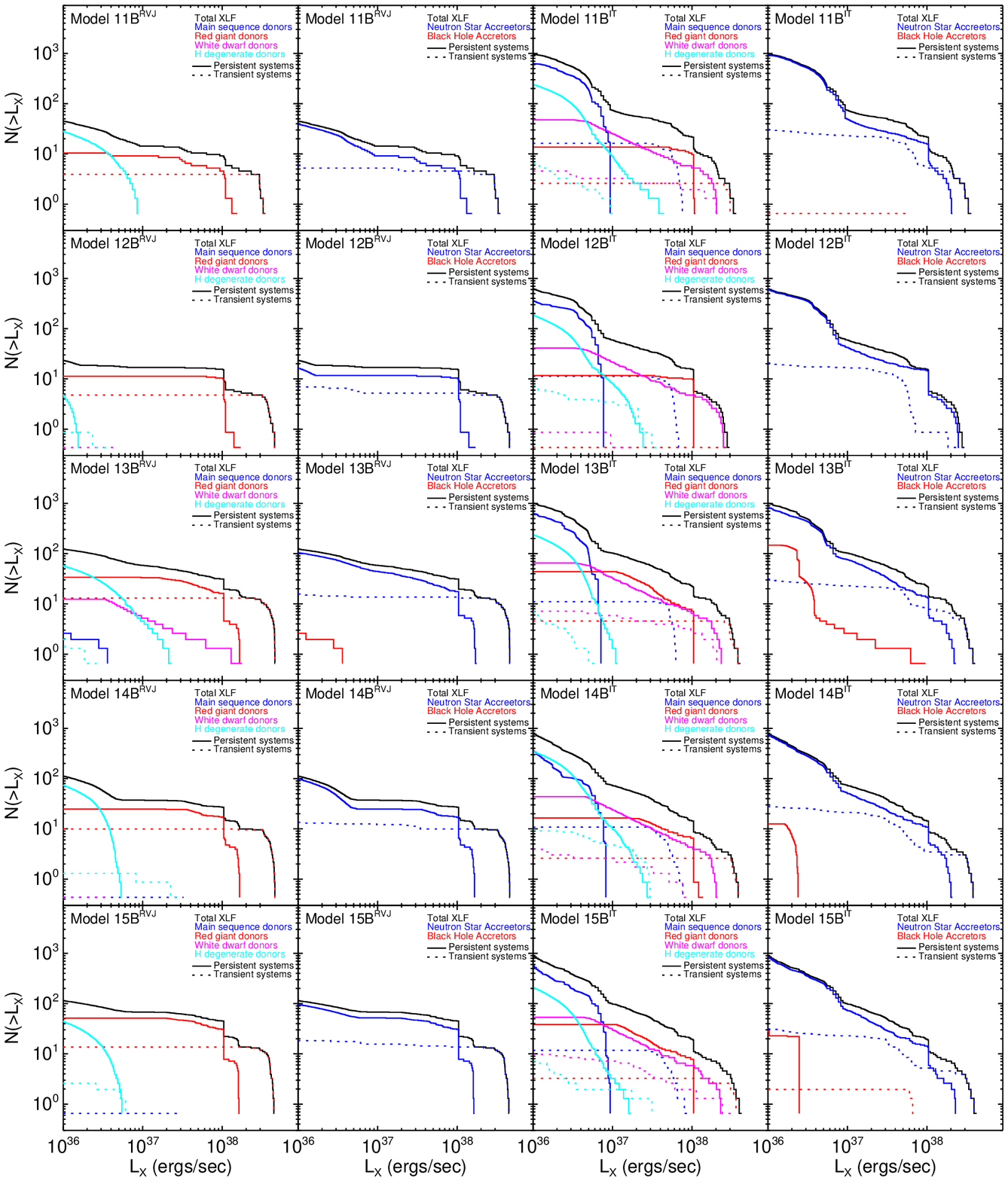}
\end{figure}

\begin{figure}
\plotone{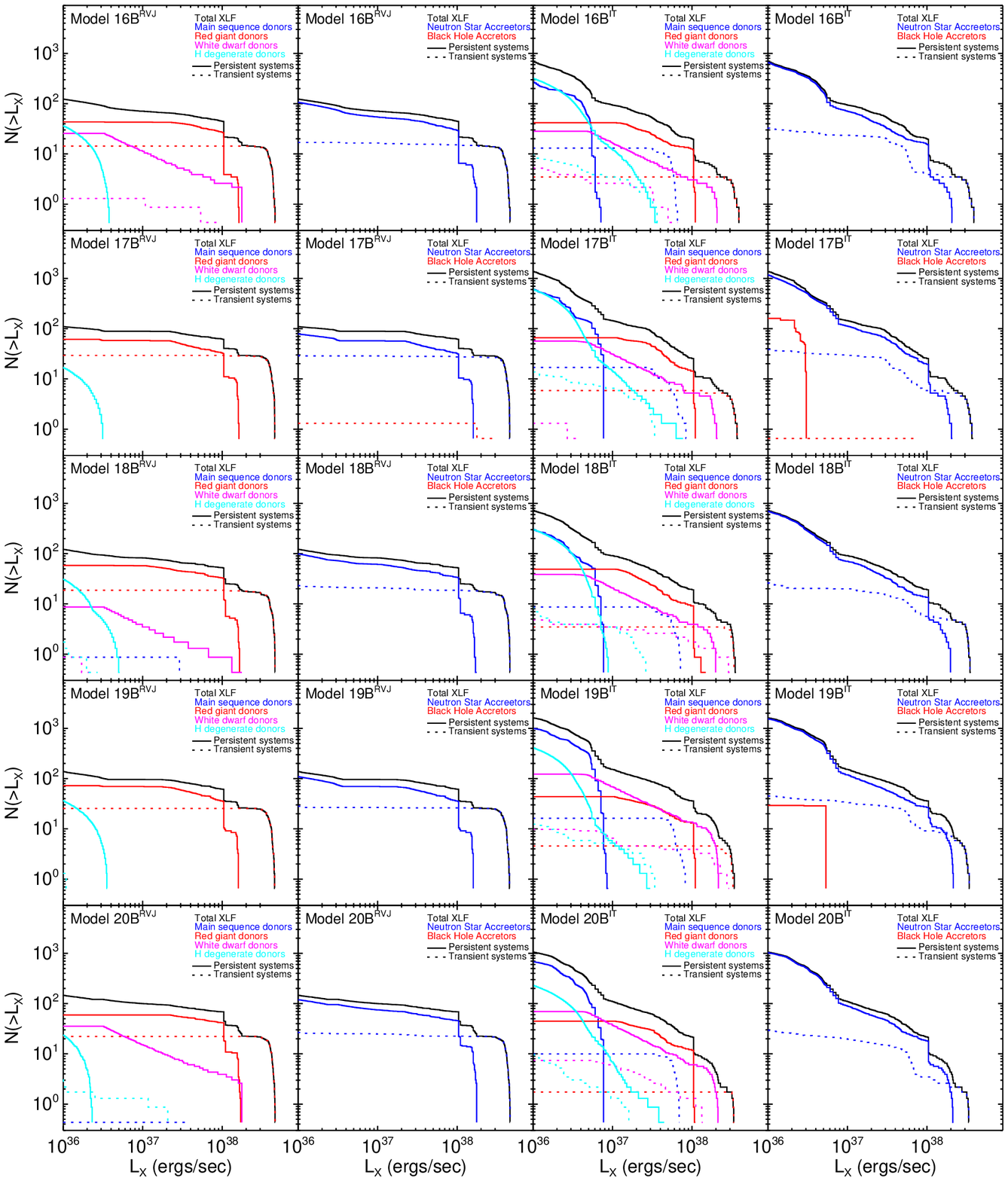}
\end{figure}

\begin{figure}
\plotone{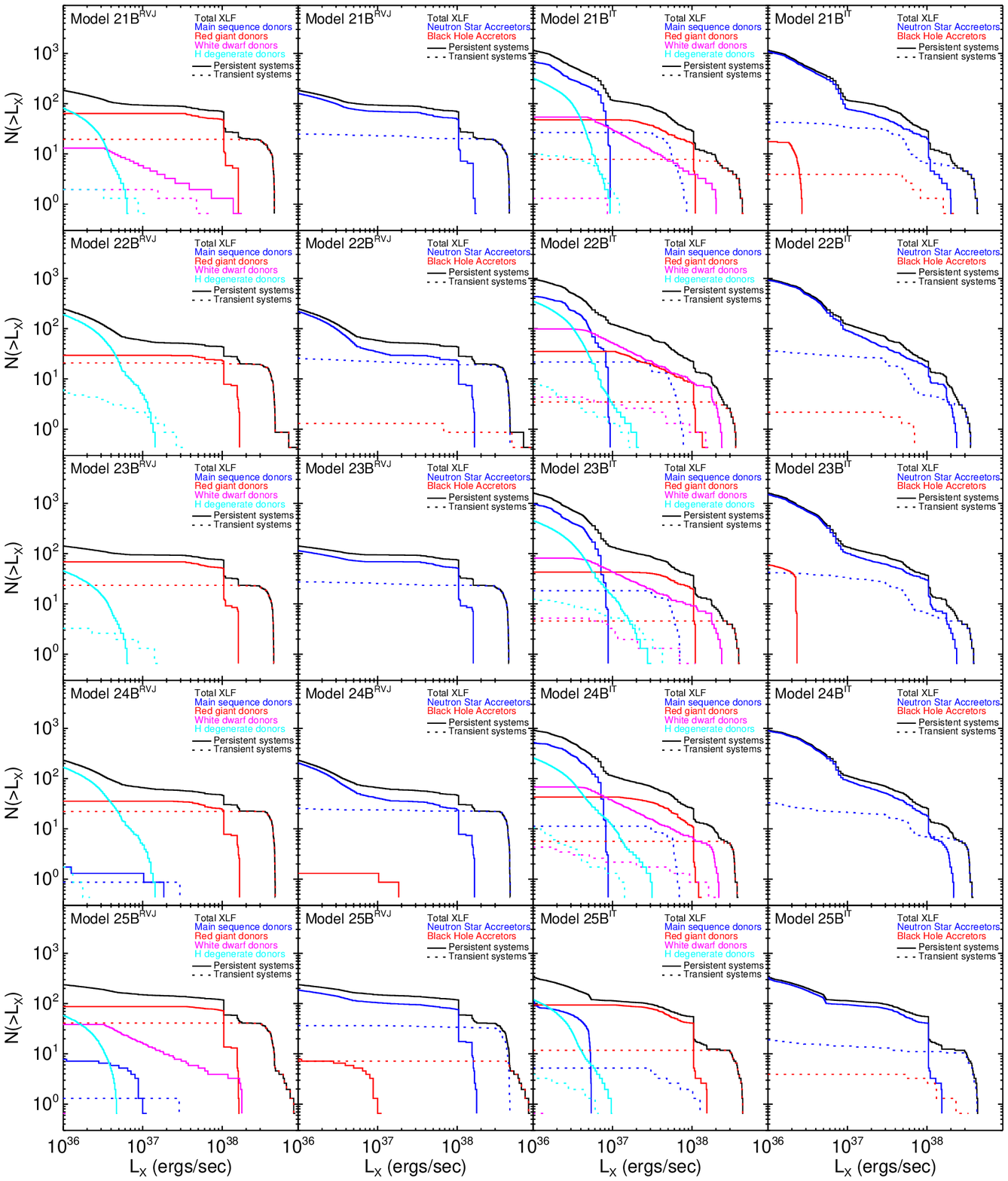}
\end{figure}

\begin{figure}
\plotone{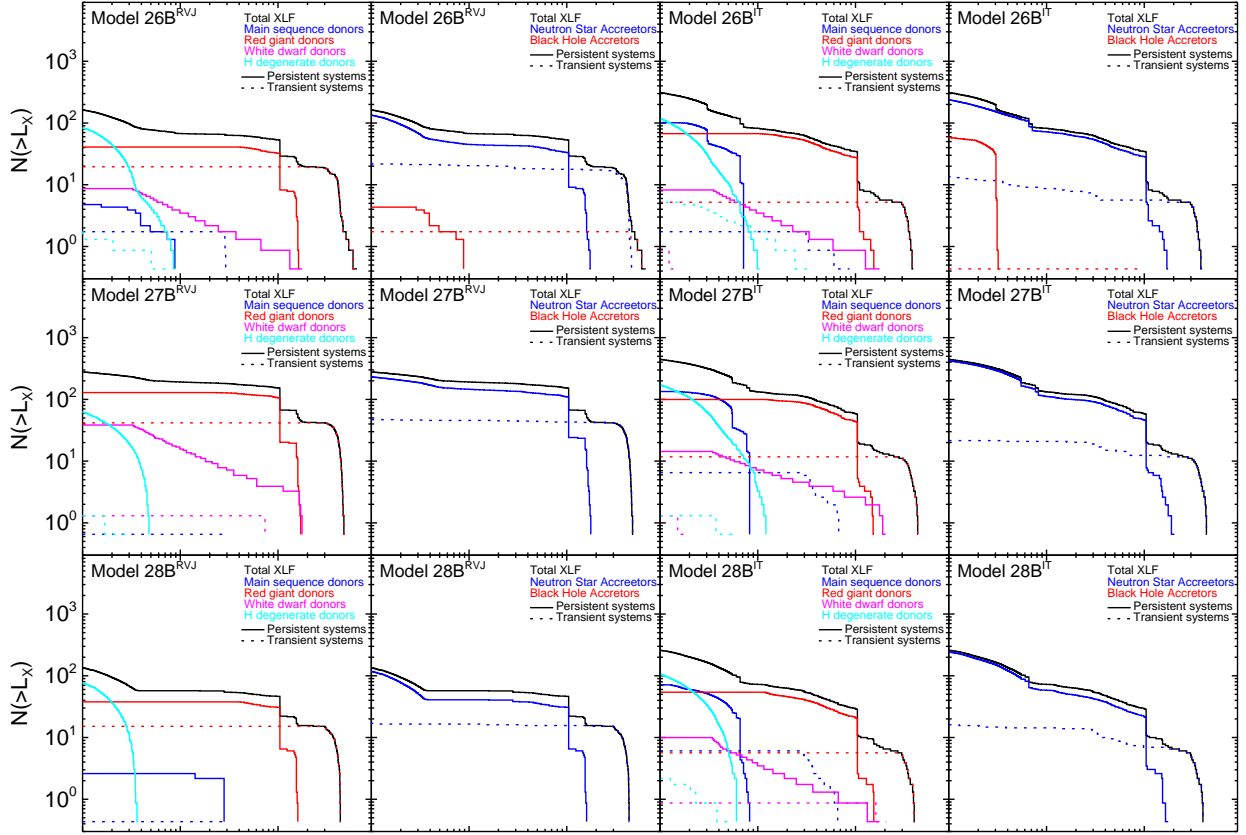}
\caption{Analysis of the LMXB population for complete list of our models. We show the contribution of different sub-populations by separating the LMXBs into groups of systems with different donor  and accretor  stellar types.}
\label{xlf_type_multi}
\end{figure}

\clearpage
\newpage

\bibliographystyle{abbrvnat}
\bibliography{ms}

\end{document}